\documentclass{kluwer}  
\usepackage{epsfig}
\newcommand{\Hm}{\rm{H}^{-}}
\newcommand{\Hp}{\rm{H}^{+}}
\newcommand{\me}{\rm{e}^{-}}
\newcommand{\mH}{\rm{H}}
\newcommand{\HD}{\rm{HD}}
\newcommand{\mHt}{\rm{H}_{2}}
\newcommand{\mHtp}{\rm{H}_{2}^{+}}
\newcommand{\hi}{\hbox{H\,{\sc i}}\,}
\newcommand{\hii}{\hbox{H\,{\sc ii}}\,}
\newcommand{\heii}{\hbox{He\,{\sc ii}}\,}
\newcommand{\heiii}{\hbox{He\,{\sc iii}}\,}
\newcommand{\citet}[1]{\citeauthor{#1} (\citeyear{#1})}
\newcommand{\citealt}[1]{\citeauthor{#1}, \citeyear{#1}}
\newcommand{\msun}{{\rm M}_{\odot}}
\newcommand{\lsun}{{\rm L}_{\odot}}
\newcommand{\rsun}{{\rm R}_{\odot}}
\newcommand{\apj}{{\em Astrophys.\ J.}}
\newcommand{\apjs}{{\em Astrophys.\ J.\ Suppl.}}
\newcommand{\mnras}{{\em Monthly Not.\ Roy.\ Astron.\ Soc.}}
\newcommand{\pasj}{{\em Pub.\ Astron.\ Soc.\ Japan}}
\newcounter{saveqn}
\newcounter{reaction}
\newcommand{\dd}[1]{{\rm d}#1}
\def\simless{\mathbin{\lower 3pt\hbox
   {$\rlap{\raise 5pt\hbox{$\char'074$}}\mathchar"7218$}}}   
\def\simgreat{\mathbin{\lower 3pt\hbox  
   {$\rlap{\raise 5pt\hbox{$\char'076$}}\mathchar"7218$}}} 

\begin{document}                       
\begin{article}
\begin{opening}         
\title{The Formation of the First Stars in the Universe} 
\author{Simon \surname{Glover}}  
\runningauthor{Simon Glover}
\runningtitle{Formation of the First Stars}
\institute{Department of Astrophysics, American Museum of Natural History, \\
       Central Park West at 79th Street, New York, NY 10024}

\begin{abstract}
In this review, I survey our current understanding of how the very first
stars in the universe formed, with a focus on three main areas of interest: 
the formation of the first protogalaxies and the cooling of gas within them,
the nature and extent of fragmentation within the cool gas, and the physics
-- in particular the interplay between protostellar accretion and protostellar
feedback -- that serves to determine the final stellar mass.

In each of these areas, I have attempted to show how our thinking has 
developed over recent years, aided in large part by the increasing ease
with which we can now perform detailed numerical simulations of primordial 
star formation. I have also tried to indicate the areas where our 
understanding remains incomplete, and to identify some of the most important
unsolved problems.
\end{abstract}

\end{opening}           

\section{Introduction}  
For more than a decade -- ever since the first release of the {\sc cobe} 
results \cite{cobe,cobe2} -- astrophysicists and cosmologists have found 
themselves in the unusual situation of knowing more about the state of the 
universe when it was only 380,000 years old than when it was 200 million 
years old. Increasingly precise measurements of the cosmic microwave 
background (CMB), as exemplified by the recent results from {\sc wmap} 
\cite{wmap}, together with a broad range of other observational constraints
\cite{sn98,sn99a,sn99b,vst00,free01,2df01,deut01,deut03} have helped
to confirm that we live in a flat universe, with approximately 5\% of the
closure density being provided by baryons, 25\% by cold dark matter (CDM),
and the remaining 70\% by some form of `dark energy' or cosmological
constant. Models of such a universe -- generally known as $\Lambda$CDM 
models -- have been heavily studied for a number of years (see, e.g.\
\citealt{ss91}; \citealt{cpt}; \citealt{gn96a}ab) and many of their 
features are well understood. For instance, the evolution
of the small inhomogeneities in the early universe that give rise to the
observed temperature anisotropies in the CMB can be followed in great detail
\cite{cmbfast}, and the resulting predictions have been strongly confirmed 
by the {\sc wmap} results. 

The evolution of the dark matter component of the universe subsequent to the
epoch of last scattering at $z \simeq 1100$ has also been studied 
intensively, using a wide range of techniques (see, for instance 
\citealt{seljak00}; \citealt{benson01}; \citealt{cooray02}). The general
agreement between the results of these studies and an increasing 
number of observational tests (e.g.\ \citealt{gray02}) has lent further 
support to this overall picture, although some puzzles remain 
\cite{moore,navarro}.

When it comes to understanding the behaviour of the baryonic component,
however, we are on much shakier ground. Although cosmological perturbation 
theory has given us a fairly good understanding of the behaviour of the 
baryons in the linear regime \cite{gh98,meik99,sm02}, many details of the 
non-linear evolution of the baryons and the development of stars and galaxies 
are not understood. At the same time, we have little or no
observational data to guide us. Although we now have observational probes of 
the Universe at redshifts $z > 6$, thanks to the success of the Sloan Digital 
Sky Survey at finding high-redshift quasars \cite{fan03}, the strong
metal lines observed in many of these quasars \cite{fan01} are evidence that
we are not yet probing the earliest epochs of star formation.

Since observational limitations prevent us, for the time being, from directly
studying the formation of the first stars and galaxies, work in this area has
been primarily theoretical in nature. Although developing a theoretical 
understanding of primordial star formation may seem at first to be a hopelessly
optimistic ambition -- after all, there is still much that we do not understand
about local star formation, despite the large quantity of observational data 
available to us -- there are actually several good reasons to think that 
the problem may be a simpler one than understanding present day star 
formation. 

First, the initial conditions -- small perturbations to a uniform cosmological
background -- are simple and well understood (provided that the $\Lambda$CDM 
model remains as accurate on small scales as it has proved to be on the 
larger scales probed by galaxy surveys and the CMB). Second, the chemistry of
primordial gas is also simple, at least in comparison to that of present day
molecular clouds. It is therefore much easier to identify the critical 
reactions and to numerically simulate the chemical evolution of the gas.
Third, magnetic fields, if present, are unlikely to be dynamically 
significant \cite{widrow}; consequently, they are usually ignored. Finally,
by restricting our attention to the first generation of stars to form, we 
can avoid the many complications posed by the feedback of stars on their 
surroundings (see, for instance, \citealt{glover01} and references therein).

Nevertheless, the problem remains a challenging one that involves processes
occurring over a very wide range of length scales, from the cosmological to
the protostellar. A popular approach is to break this problem up into a
series of simpler problems, with different characteristic scales, that can
be tackled individually. For instance: 
\begin{enumerate}
\item[(i)] When do the first protogalaxies\footnote{A note on terminology:
in this review, I use the term `protogalaxy' as a convenient shorthand for 
`gravitationally bound gas cloud': the fact that something is described as 
a protogalaxy does not imply that it is actively forming stars, merely that
it has the potential to do so.} form, and how massive are they?
\item[(ii)] How does gas evolve within these protogalaxies? Does it 
fragment, and if so, how large are the resulting fragments?
When and why does fragmentation stop?
\item[(iii)] What is the initial mass function (IMF) of the first stars, and
what processes determine this?
\end{enumerate}
In this article, I review our progress at answering these questions. The body
of the review is divided into three sections, each with a theme that broadly 
reflects one of the questions posed above, although there is inevitably a 
certain amount of overlap.

There are, of course, many interesting questions concerning the first stars
and galaxies which I have neither the time nor the space to properly address 
in this review; for instance, the question of how best to go about observing
them; or the question of how they affect their environment both on small and
large scales. Some of these questions are addressed in other recent reviews
of early star formation \cite{bl01,blar04,cf04}, which complement the material
presented here.

Throughout this paper, unless otherwise indicated, I adopt cosmological 
parameters taken from the {\sc wmap} concordance model \cite{wmap_model}.
Specifically: $\Omega_{\rm m} = 0.29\, $, $\Omega_{\Lambda} = 0.71\, $, 
$\Omega_{\rm b} = 0.047\, $, $h = 0.72\, $, $\sigma_{8} = 0.9\, $, 
$n_{\rm s} = 0.99\, $.

\section{The formation of protogalaxies}
\label{proto}
\subsection{The first bound objects}
\label{first-obj}
In CDM models, gravitationally bound objects form in a hierarchical, or 
`bottom-up' fashion, with the smallest, least massive objects forming first,
and larger objects forming later through a mixture of mergers and accretion.
The mass scale on which gravitationally bound objects begin to form 
(i.e.\ the minimum mass of a bound object) is set by the free streaming
of the dark matter particles \cite{blum84}. In general, this mass 
scale is many orders of magnitude smaller than scales of cosmological 
interest; for instance, in the neutralino model for CDM, $M_{\rm min} 
\simeq 10^{-7} \: \msun$ \cite{hof01}. However, the subsequent formation
of larger objects occurs rapidly, and at most redshifts a large number
of gravitationally bound objects (frequently referred to as `dark matter 
halos') exist, with a wide range of different masses. 

Considerable effort has been devoted to determining the mass function of
dark matter halos as a function of redshift. The most widely used expression
for the mass function is one originally suggested by \citet{ps}:
\begin{equation}
 \label{ps_mass}
  n(M,z)\, \dd{M} = \sqrt{\frac{2}{\pi}} 
 \frac{\rho_{\rm dm}}{M} \frac{\dd{\nu}}{\dd{M}} 
 \exp \left( - \frac{\nu^{2}}{2} \right) \dd{M}.
\end{equation}
Here $n(M,z)\, \dd{M}$ is the comoving number density of halos at redshift 
$z$ with dark matter masses in the interval $(M, M + \dd{M})$, $\rho_{dm}$ 
is the cosmological background density of dark matter, and 
$\nu \equiv \delta_{\rm c}/[D(z) \sigma(M)]$, where 
$\delta_{\rm c}$ is a critical overdensity (generally taken to be 1.69), 
$D(z)$ is the linear growth factor \cite{peeb80,cpt} and $\sigma(M)$ 
is the rms fluctuation in the cosmological density field of dark matter 
smoothed on a mass scale $M$. A comprehensive discussion of the derivation 
of this equation is given in \citet{bcek}.

Comparisons with the results of N-body simulations at low redshift 
\cite{jenk01} and at high redshift \cite{hjc} demonstrate that the 
Press-Schechter mass function provides a reasonable fit to the true 
mass function, although a better fit to the simulation results can be 
obtained by using a modified form suggested by \citet{st99}:
\begin{equation}
 \label{mod_ps_mass}
  n(M,z)\, \dd{M} = A \left(1 + \frac{1}{\nu'^{2q}}\right) 
 \sqrt{\frac{2}{\pi}} \frac{\rho_{\rm dm}}{M} 
 \frac{\dd{\nu'}}{\dd{M}} \exp \left( - \frac{\nu'^{2}}{2} \right) \dd{M},
\end{equation}
where $\nu' = \sqrt{a}\nu$, $a=0.707$, $A \simeq 0.322$ and $q=0.3$.

In either case, the basic form of the mass function is the same: it behaves
as a power law for $\nu \ll 1$ and falls off exponentially for $\nu \gg 1$.
In CDM models, $\sigma(M)$ decreases monotonically with increasing mass,
and so the most massive objects will also be the rarest. The transition to
exponential behaviour occurs for $\nu \sim 1$, or $\sigma(M) \sim 
\delta_{\rm c} / D(z)$, and so this transition occurs at a progressively 
smaller mass as we move to higher redshifts.

Given a mass function of this type, is there any way to specify when the 
first halo of a given mass forms? Strictly speaking, the answer is no; the
probability of finding a halo of any finite mass is never zero. In practice,
however, we are more interested in determining when this probability grows to
some interesting size, or when the number density of halos exceeds some
specified threshold (which amounts to the same thing). This is most commonly
done by specifying a value of $\nu$ which is of interest; for instance, 
reference is often made to $3\sigma$ halos, which are simply halos for
which $\nu = 3$ and which therefore have a dark matter mass $M$ satisfying:
\begin{equation}
 \sigma(M) = \frac{1}{3} \frac{\delta_{\rm c}}{D(z)}.  \label{sigma}
\end{equation} 
Such halos are moderately rare objects, representing no more than a few 
thousandths of the total cosmic mass \cite{mw02}, but are sufficiently common
that one would expect to find many of them within a single Hubble volume. 
They are often taken to be representative of the earliest objects to form, 
although this choice is somewhat arbitrary. 

Unfortunately, while the Press-Schechter approach allows us to determine 
when the first dark matter halos of a given mass form, it does not, by itself,
tell us when the first {\em protogalaxies} form, as it contains no information
about the behaviour of the baryonic component of the universe. 
Unlike the dark matter, the baryons do not initially form structures 
on very small scales, since pressure forces act to suppress the growth
of small-scale perturbations \cite{jeans02,jeans28,bonnor57}. We can
estimate the scale on which pressure forces become significant by
equating the sound-crossing timescale, $t_{\rm sc}$, with the gravitational
free-fall timescale, $t_{\rm ff}$: if $t_{\rm sc} < t_{\rm ff}$ then 
perturbations can respond subsonically to changes in the gravitational field 
and will therefore remain in approximate hydrostatic equilibrium; on the other 
hand, if $t_{\rm sc} > t_{\rm ff}$ then perturbations cannot
respond subsonically, and some degree of gravitational collapse becomes 
inevitable. In gas with a density $\rho$ and sound speed $c_{\rm s}$, we 
would therefore expect collapse to be suppressed on scales 
\begin{equation}
 \lambda \simless \frac{c_{\rm s}}{\sqrt{G\rho}}.
\end{equation}
A more careful analysis using linear perturbation theory \cite{peeb80}
shows that in a purely baryonic universe, the growth of perturbations
is completely suppressed on scales smaller than
\begin{equation}
\label{cosmic_lj}
 \lambda_{\rm J} \leq \frac{\pi^{1/2}c_{\rm s}}{\sqrt{G\rho_{\rm b}}},
\end{equation}
where $\rho_{\rm b}$ is the cosmological baryon density. This critical 
wavelength is commonly known as the Jeans length. The associated mass scale, 
known as the Jeans mass, is conventionally defined as 
\begin{equation}
 M_{\rm J} = \frac{4}{3} \pi \rho_{\rm b}   
 \left(\frac{\lambda_{\rm J}}{2}\right)^{3}.
\end{equation}

The value of the Jeans mass depends on the baryon density, which is a 
simple function of redshift, and on the temperature of the intergalactic 
medium (through the dependence of $\lambda_{\rm J}$ on $c_{\rm s}$). The 
latter is simple to calculate at epochs prior to the onset of  widespread 
star formation and is well approximated by \cite{gp}
\begin{equation} 
 T = 410 \left(\frac{1+z}{150}\right)^{2} \: {\rm K}
\end{equation} 
for redshifts $z < 150$. The corresponding Jeans mass at these redshifts 
is given by
\begin{equation}
  M_{\rm J} = \frac{4.9 \times 10^{4}}{(\Omega_{\rm b} h^{2})^{1/2}} 
\left(\frac{1+z}{150}\right)^{3/2} \: \msun.
\end{equation} 

To generalize this to the case of a universe containing both baryons and 
cold dark matter, it is tempting to simply replace the baryon density in 
the above equations with the total density $\rho_{\rm m} = \rho_{\rm b} + 
\rho_{\rm dm}$, which would give us
\begin{equation}
 \label{mj}
  M_{\rm J} = \frac{4.9 \times 10^{4}}{(\Omega_{\rm m} h^{2})^{1/2}} 
\left(\frac{1+z}{150}\right)^{3/2} \: \msun
\end{equation}
for $z < 150$; or in other words, a Jeans mass that is a factor 
$(\Omega_{\rm b} / \Omega_{\rm m})^{1/2}$ smaller. In fact, the situation
is not so simple, as perturbations can continue to grow on small scales 
in the dark matter even when suppressed in the baryons. A linear treatment 
of this case is given in \citet{gh98}, but ultimately this treatment
breaks down as small-scale structure in the dark matter begins to grow
non-linearly. Although these non-linear effects have received little direct 
study, there is some evidence from numerical simulations that they can cause
baryons to collapse on scales smaller than $\lambda_{\rm J}$ (see, for 
instance, the discussion in section~2.1 of \citealt{hl97}), although any such
collapse will be significantly delayed relative to the dark matter due to the 
influence of the gas pressure. In view of this, it is probably best to treat 
the value of $M_{\rm J}$ given by Equation~(\ref{mj}) as an estimate of the 
scale on which pressure effects begin to dominate, rather than as an absolute 
lower limit to the protogalactic mass.

\begin{figure}
\centering
\epsfig{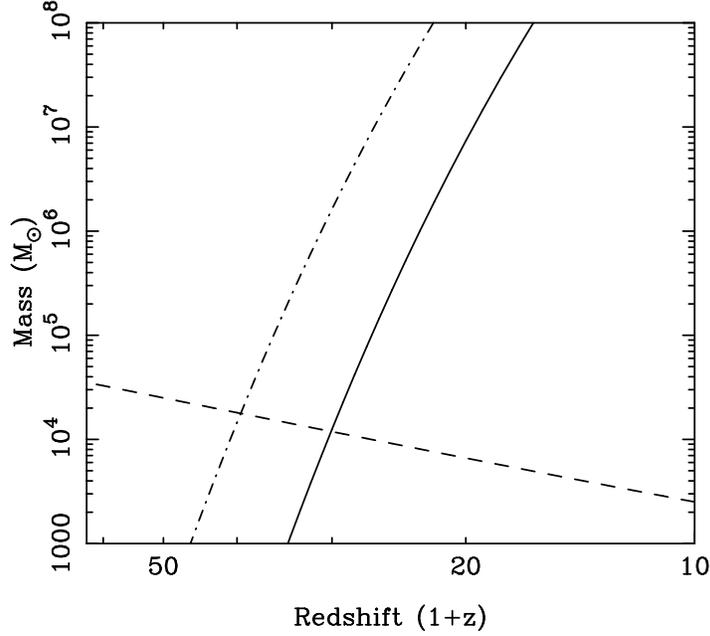}
\caption{The evolution with redshift of $M_{3\sigma}$ (solid line), 
$M_{4\sigma}$ (dot-dashed line) and $M_{\rm J}$ (dashed line). 
Protogalaxies will develop within $3\sigma$ dark matter halos once 
the mass of dark matter in the halo, $M_{3\sigma}$, exceeds $M_{\rm J}$;
this occurs at $z \simless 30$. Similarly, protogalaxies will form in 
$4\sigma$ halos once their dark matter mass, $M_{4\sigma}$, exceeds 
$M_{\rm J}$, which occurs at $z \simless 40$.}
\label{mass_comp}
\end{figure}
Given $M_{\rm J}$, we can go on to estimate the mass and formation redshift
of the first protogalaxies by asking when the total mass of a $3\sigma$ halo
first exceeds the Jeans mass. To properly answer this question, we would need
to know the baryon fraction of these protogalaxies (i.e.\ their ratio of 
baryonic to dark matter). In practice, however, we know that this will be 
small and that the protogalactic mass will be dominated by the dark matter 
component. Therefore, for the purposes of a simple estimate it is sufficient
to compare the Jeans mass with the dark matter mass of the $3\sigma$ halo
(hereafter $M_{3\sigma}$), which we can calculate using Equation~(\ref{sigma}).
The evolution with redshift of both mass scales is plotted in 
figure~\ref{mass_comp}. We can see from the figure that the first 
protogalaxies will have a total mass $M \sim 10^{4} \: \msun$ and will form at 
a redshift $z \sim 30$. It is also clear that uncertainties 
in $M_{\rm J}$ will have little effect on the estimated redshift, due to the 
sharp rise in $M_{3\sigma}$ with declining $z$. On the other hand, the use 
of a different criterion to identify our `first' objects (e.g.\ considering 
$4\sigma$ halos instead of $3\sigma$ ones) has a rather larger effect on  $z$,
but has very little effect on the estimated protogalactic mass.

\subsection{The importance of cooling} 
\label{imp_cool}
Once a protogalaxy has formed, the next task is to determine how the gas
within it evolves. In particular, we would like to know whether {\em every} 
protogalaxy that forms is capable of forming stars, or whether there are
other prerequisites.

We can gain considerable insight into this question by considering the
thermal evolution of a parcel of gas that is undergoing gravitational 
collapse. The gravitational potential energy of the gas is transformed
first into kinetic energy and thence into thermal energy through adiabatic
compression, as well as the action of shocks if the flow is supersonic. Unless
the gas can dissipate this thermal energy through radiative cooling, it
must inevitably heat up. Since both density and temperature are rising,
the pressure will increase rapidly, and ultimately will become large enough
to halt the collapse. 

We can make this argument more quantitative by considering the gravitational 
stability of small perturbations within the collapsing gas. As in the
cosmological case, we can derive a minimum unstable mass scale, again
termed the Jeans mass, which scales as
\begin{equation}
 M_{\rm J} \propto \frac{c_{\rm s}^{3}}{\rho^{1/2}} \propto 
 \frac{T^{3/2}}{\rho^{1/2}}.
\end{equation}
If we define an effective adiabatic index 
\begin{equation}
 \gamma_{\rm eff} = 1 + \frac{\dd{{\rm ln}\, T}}{\dd{{\rm ln}\, \rho}},
\end{equation}
then $M_{\rm J}$ will evolve with density as
\begin{equation}
 M_{\rm J} \propto \rho^{\frac{3}{2}(\gamma_{\rm eff} - \frac{4}{3})}.
\end{equation}
Therefore, if $\gamma_{\rm eff} > 4/3$, the Jeans mass will increase
during the collapse and will eventually become comparable to the mass
of the protogalaxy, at which point collapse must halt. Since
$\gamma_{\rm eff} = 5/3$ for an atomic gas evolving adiabatically, it is 
clear that in the absence of radiative cooling, the increasing 
thermal pressure will bring collapse to an end long before protostellar 
densities  are reached. Therefore, star formation is only possible if the 
gas can cool.

The timescale on which cooling occurs is also of great importance. It has 
long been argued \cite{gt76,ro,silk77a} that the behaviour of gas in a 
collapsing protogalaxy depends upon the relative sizes of its cooling
timescale,
\begin{equation}
 t_{\rm cool} = \frac{1}{\gamma - 1}\frac{nkT}{\Lambda(T)},
\end{equation}
(where $n$ is the particle number density and $\Lambda(T)$ is the cooling 
rate per unit volume), its dynamical (or free-fall) timescale, given by 
\begin{equation}
 t_{\rm dyn} = \sqrt{\frac{3}{32 \pi G \rho}},
\end{equation}
and the cosmological timescale, or Hubble time, 
\begin{equation}
 t_{\rm H} \simeq \frac{1}{H(z)}.
\end{equation}
It is easy to show that $ t_{\rm dyn}$ is always less than $t_{\rm H}$,
so there are only three possible arrangements:
\begin{enumerate}
\item[(i)] $\:t_{\rm cool} > t_{\rm H} > t_{\rm dyn}$,
\item[(ii)] $\:t_{\rm H} > t_{\rm cool} > t_{\rm dyn}$,
\item[(iii)] $\:t_{\rm H} > t_{\rm dyn} > t_{\rm cool}$.
\end{enumerate}

In case (i), cooling takes place on a cosmological timescale, and is so slow
that the gas evolves much as if there were no cooling at all. It quickly
becomes pressure supported and remains so almost indefinitely, unless 
disturbed by an external event, such as a merger with another protogalaxy.
In case (ii), the gas also becomes pressure supported, but subsequently
contracts quasi-statically on a cosmologically interesting timescale. Finally,
gas described by case (iii) never becomes pressure supported, but instead
simply collapses at or near the free-fall rate.

In practice, the situation is often far more complex than this analysis 
suggests, since the appropriate description for gas in a given protogalaxy
may vary with its location within the protogalaxy, and may also change over 
time as the density, temperature and/or chemical makeup of the gas change.
Nevertheless, this scheme is a useful first approximation, and serves to 
further highlight the central role played by radiative cooling.

\subsection{Cooling and chemistry within primordial gas}
A number of potential cooling mechanisms exist within primordial gas
\cite{aanz97b}, but many, such as Lyman-$\alpha$ cooling, operate only
for $T > 10^{4} \: {\rm K}$, while the first protogalaxies have 
characteristic temperatures $T \sim 100$--$1000 \: \rm{K}$. At these low 
temperatures, the dominant coolant is molecular hydrogen, $\mHt$, the
most abundant primordial molecule. Therefore, to determine the cooling rate 
accurately, we require an accurate value for the $\mHt$ abundance, which 
means that in addition to studying the thermal evolution of the gas we must
also study its chemical evolution. 

The chemistry of primordial gas
has been investigated by a number of authors (\citealt{dl87}; \citealt{bl91}; 
\citealt{aanz97a}; \citealt{gp}; \citeauthor{sld98}, \citeyear{sld98}, 
\citeyear{sld02}) and proves to be surprisingly complex despite the 
limited number of elements involved. This complexity is due to the wide
variety of different molecules and molecular ions that can be formed. 
However, if we are only interested in those aspects of the chemistry that 
affect the cooling rate, then we can make substantial simplifications 
\cite{aanz97a}: the chemical model can be reduced to a few processes that 
determine the ionization balance of the gas (e.g.\ collisional ionization, 
radiative recombination), together with those reactions involved in the 
formation and destruction of $\mHt$. 

The formation of $\mHt$ in local molecular clouds occurs primarily on the 
surface of interstellar dust grains: hydrogen atoms are adsorbed onto the 
surface of the grains, react to form $\mHt$ and subsequently escape back 
into the interstellar medium \cite{gs63}. In primordial gas, however, there is 
no dust, and so no possibility of forming $\mHt$ by this process. Instead, 
$\mHt$ formation is dominated by various sets of gas phase reactions. 

The simplest gas-phase reaction -- direct radiative association of two
hydrogen atoms to form $\mHt$:
\setcounter{saveqn}{\value{equation}}
\setcounter{equation}{0}
\renewcommand{\theequation}{\mbox{R\arabic{equation}}}
\begin{equation}
 \mH + \mH \rightarrow \mHt + \gamma,
\end{equation}
is strongly forbidden unless one of the hydrogen atoms is in an excited 
electronic state, and therefore plays an important role only in rather
unusual circumstances, such as in the intergalactic medium near the end
of the epoch of recombination \cite{lb91,rdb93}. It does not significantly
influence protogalactic $\mHt$ formation.

Three-body formation of $\mHt$, via the reactions 
\begin{eqnarray}
 \mH + \mH + \mH  & \rightarrow & \mHt + \mH,  \label{h23a} \\
 \mH + \mH + \mHt & \rightarrow & \mHt + \mHt, \label{h23b}
\end{eqnarray}
{\em can} play a significant role \cite{pss83}, but only at high densities 
($n_{\mH} \simgreat 10^{8} \: {\rm cm}^{-3}$), since the rate coefficients
of these reactions are small. At lower densities, gas-phase formation of 
$\mHt$ is dominated by two sets of reactions. The first involves the $\Hm$ 
ion as an intermediate state
\begin{eqnarray}
 \mH + \me & \rightarrow & \Hm + \gamma,  \label{h2f1} \\
 \Hm + \mH & \rightarrow & \mHt + \me,   \label{h2f2}
\end{eqnarray}
and was first discussed in the context of the local ISM by \citet{mc61}, and
in a cosmological context by \citet{pd68}. The second set of reactions involves
the $\mHtp$ ion as an intermediary, and was first discussed in a cosmological 
context by \citet{sz67}
\begin{eqnarray}
 \mH + \Hp & \rightarrow & \mHtp + \gamma, \label{h2f3} \\
 \mHtp + \mH & \rightarrow & \mHt + \Hp. \label{h2f4}
\end{eqnarray}
\setcounter{reaction}{\value{equation}}
\setcounter{equation}{\value{saveqn}}
\renewcommand{\theequation}{\arabic{equation}}
These two sets of reactions (hereafter the $\Hm$ pathway and the $\mHtp$ 
pathway respectively) share two important characteristics. Firstly, both
are limited by their initial step, since the radiative association reactions 
\ref{h2f1} and \ref{h2f3} occur at a much slower rate than the subsequent 
ion-neutral reactions \ref{h2f2} and \ref{h2f4}. Secondly, the role played
by free electrons in the $\Hm$ pathway is extremely similar to the role played
by $\Hp$ ions in the $\mHtp$ pathway, and in both cases the $\mHt$ formation
rate is directly proportional to the fractional ionization of the gas, 
provided that the latter is small.\footnote{If the fractional ionization is 
large, then the mutual neutralization of $\Hm$ with $\Hp$ and the dissociative
recombination of $\mHtp$ become significant, and this simple relationship 
breaks down; this is discussed in more detail in \citet{glover03}.}

The main difference between the two pathways stems from the difference
in the rates of reactions \ref{h2f1} and \ref{h2f3}: $\Hm$ forms via 
\ref{h2f1} much faster than $\mHtp$ forms via \ref{h2f3}, and so the 
$\Hm$ pathway generally dominates the gas phase production of $\mHt$. 

The dependence of $\mHt$ formation on the presence of free electrons and 
protons might lead one to suppose that $\mHt$ formation will be very 
inefficient in low temperature gas, since the equilibrium ionization fraction
is very low. In practice, however, moderate amounts of $\mHt$ can be formed
if the protogalactic gas is not initially in ionization equilibrium. This
is certainly the case in newly-formed protogalaxies, since the IGM itself
is not in ionization equilibrium; instead, it retains a residual fractional 
ionization dating from the epoch of recombination. This comes about because
the Hubble expansion ensures that the recombination timescale 
exceeds the expansion timescale before the IGM can reach equilibrium, 
freezing the fractional ionization at a value of approximately 
$2 \times 10^{-4}$ \cite{sld98}. Protogalaxies forming from the IGM therefore
begin with this small non-equilibrium fractional ionization. 

Once a protogalaxy has formed, this residual ionization quickly vanishes,
as the increased density leads to a greatly increased recombination rate. 
However, there remains a brief window of opportunity in which $\mHt$ can form. 
Simple estimates of the resulting molecular fraction have been given by a 
number of authors \cite{suny98,ns99,oh02} and are typically in the range 
$f_{\mHt} = 10^{-3}$--$10^{-4}$. For comparison, note that the molecular 
fraction in the IGM at this time is approximately $2 \times 10^{-6}$ \cite{gp}.

\begin{figure}
\centering
\epsfig{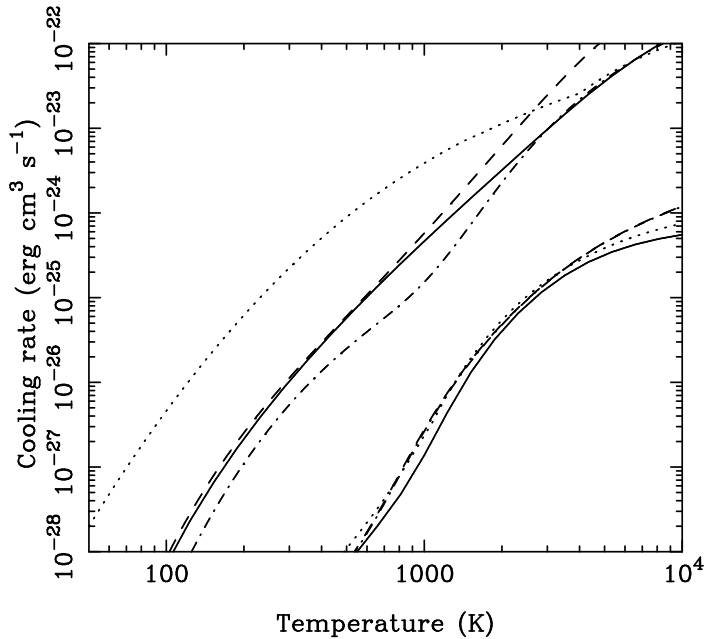}
\caption{A comparison of various parameterizations of the $\mHt$ cooling 
function, plotted in units of ${\rm erg} \: {\rm cm}^{3} \: {\rm s}^{-1}$.
Rates are computed assuming that $n_{\mH} \gg n_{\mHt}$, and that the
ortho-to-para ratio is 3:1. The lower set of lines corresponds to a gas 
density $n_{\mH} = 10^{6} \: {\rm cm}^{-3}$; the upper set corresponds 
to $n_{\mH} = 10^{0} \: {\rm cm}^{-3}$. Solid line -- 
Le~Bourlot, Pineau~des~For\^{e}ts, and Flower~(1999); 
dashed line -- Galli and Palla~(1998); dotted line -- Lepp and Shull~(1983); 
dash-dotted line -- Hollenbach and McKee~(1979).}
\label{h2cool}
\end{figure}

Given the $\mHt$ abundance, density and temperature, it is then a simple
matter to calculate the $\mHt$ cooling rate. Various parameterizations of
this rate have been given in the literature; figure~\ref{h2cool} shows some 
commonly cited examples, plotted as $\Lambda_{\mHt} (n_{\mH} n_{\mHt})^{-1} $, 
where $\Lambda_{\mHt}$ is the $\mHt$ cooling rate per unit volume. The basic 
features of the cooling rate are straightforward: it falls off exponentially 
at low temperatures, due to the rather large excitation energy of the first 
accessible excited state (the $J=2$ rotational state, which lies 512K above 
the $J=0$ para-hydrogen ground state), and is essentially negligible 
below $100 \: {\rm K}$; 
it scales with density as $\Lambda_{\mHt} \propto n_{\mHt}^{2}$ at low 
densities, where radiative de-excitation dominates, and as $\Lambda_{\mHt} 
\propto n_{\mHt}$ at high 
densities, where collisional de-excitation dominates and the level 
populations approach their local thermodynamic equilibrium (LTE) values. 
The transition between low density and high density behaviour occurs near 
a critical density $n_{\rm cr} \simeq 10^{4} \: {\rm cm}^{-3}$.

The major uncertainty in the determination of the $\mHt$ cooling rate comes
from uncertainties in the values of the collisional de-excitation rates, 
which are highly sensitive to the details of the potential energy surface
used to calculate them, as well as to the method of calculation adopted 
\cite{lbd95}. As a result, there exists substantial disagreement in the 
literature on the form and magnitude of the $\mHt$ cooling rate at low 
densities, as can be seen from figure~\ref{h2cool}. However, recent 
calculations have removed much of this uncertainty, with the calculated 
collisional rates having more or less converged. 

\subsection{Thermal evolution: simple models}
\label{simple}
Armed with an appropriate set of chemical reaction rates and an accurate 
$\mHt$ cooling rate (see, for example, \citealt{aanz97a} or 
\citealt{glover01}), the next step is to follow the coupled chemical, thermal 
and dynamical evolution of a protogalaxy as it forms in order to determine 
its fate. 

The simplest approach to this problem dispenses entirely with any attempt to
accurately simulate the dynamical evolution of the protogalaxy. Instead, the 
density evolution is specified in advance, and the model focuses on 
determining the chemical and thermal evolution of the gas. For instance, it
is frequently assumed that if cooling is effective, then the density 
evolution will be the same as in pressure-free collapse.
This approximation relies on the assumption that pressure gradients are
everywhere small compared to gravitational forces.

A number of authors have considered the problem of protogalactic collapse
within this framework (\citealt{mst69}; \citealt{hutch76}; 
\citeauthor{ys79}~\citeyear{ys79,ys80}; \citealt{carl81}; \citealt{pss83}; 
\citealt{vb87}; \citealt{sun96}; \citealt{om2000}; \citealt{fp01}), 
often supplementing it with the additional 
assumptions of spherical symmetry and uniform density. Use of these 
approximations reduces the problem to one of computing the chemical and
thermal evolution of a single representative parcel of gas. 

Most of these models predict the same general type of behaviour. Initially, 
the $\mHt$ cooling 
rate is negligible and the evolution of the gas is very close to adiabatic. 
As the collapse proceeds, however, the increasing temperature, density and 
$\mHt$ abundance all combine to  dramatically increase the $\mHt$ cooling 
rate and decrease $t_{\rm cool}$. Eventually, $t_{\rm cool}$ becomes 
comparable to the collapse timescale, and the collapse ceases to be even 
approximately adiabatic. Instead, the temperature reaches a peak and then 
decreases at higher densities as radiative cooling becomes increasingly 
dominant over compressional heating. The quantitative details, such as the 
value of the peak temperature, are sensitive to the treatment of $\mHt$ 
cooling and gas chemistry adopted, and generally vary from model to model, 
although never by more than a factor of a few.

The only case in which this type of model predicts substantially different
behaviour is when some other process, such as UV photodissociation, acts to
reduce the $\mHt$ abundance (see, e.g.\ \citealt{om2001}). In this case,
the protogalaxy may be unable to form sufficient $\mHt$ to cool the gas 
before it reaches a temperature and density at which collisional dissociation 
of $\mHt$ becomes significant. This results in the gas temperature continuing 
to rise until the onset of Lyman-$\alpha$ cooling at a temperature of 
approximately $10^{4} \: {\rm K}$. 

An alternative model is presented by \citet{teg97}. They make a similar set
of approximations (spherical symmetry, uniform density, free-fall collapse),
but halt the collapse when one of two conditions is met:
\begin{enumerate}
\item[(i)] The gas temperature exceeds the virial temperature of the 
protogalaxy, defined as \cite{b92}:
\begin{equation}
 \label{tvir}
 T_{\rm vir} = \frac{G M \mu m_{\rm H}}{2kR_{\rm vir}}.
\end{equation}

\item[(ii)] The mean density of the gas exceeds the mean density of the dark
matter halo. The latter can be written as 
\begin{equation}
 \bar{\rho}_{\rm DM} = (1 + \Delta) \rho_{\rm dm}, 
\end{equation}
where $\Delta = 18 \pi^{2}$ for an Einstein-de Sitter cosmology \cite{peeb80};
analogous values for open or $\Lambda$-dominated cosmological models are 
given in \citet{bn98}.
\end{enumerate}

If condition (ii) is met, then shocks are assumed to raise the gas temperature 
instantaneously to $T_{\rm vir}$ at the end of the collapse.

\citet{teg97} make no attempt to follow the further dynamical evolution of
the gas. Instead, they hold its density constant, and
study its subsequent thermal and chemical evolution. If the gas
temperature decreases by more than 25\% during an interval corresponding
to a 25\% decrease in redshift, i.e.\ if
\begin{equation}
T(0.75z_{\rm c}) \leq 0.75 T(z_{\rm c}),
\end{equation} 
where $z_{\rm c}$ is the redshift at which the collapse terminates, 
then the protogalaxy is considered to be able to cool effectively.
Otherwise, cooling is considered to be ineffective.

By performing this analysis for protogalaxies with a wide range of masses 
and collapse redshifts, 
\citeauthor{teg97} are able to map out the region in $M$-$z_{\rm c}$ 
parameter space corresponding to protogalaxies that can cool effectively.
They find that at each redshift they can identify a minimum mass 
$M_{\rm min}$ such that protogalaxies with $M > M_{\rm min}$ cool 
effectively, while those with $M < M_{\rm min}$ do not. For instance, at 
$z = 30$, they find that $M_{\rm min} \sim 10^{6} \: \msun$, two orders of 
magnitude larger than the Jeans mass at that redshift. This result can also
be expressed in terms of a minimum virial temperature, $T_{\rm min}$, related 
to $M_{\rm min}$ through Equation~(\ref{tvir}); at $z = 30$, this is
$1000 \: {\rm K}$. While the values of $M_{\rm min}$ and 
$T_{\rm min}$ obtained in this way are somewhat sensitive to the choice of 
$\mHt$ cooling function \cite{aanz98a,glover01}, the basic behaviour remains
the same.

These two approaches -- the free-fall collapse model and the 
\citeauthor{teg97} model -- therefore present us with two distinct 
scenarios for the formation of protogalaxies. The free-fall models
predict that every protogalaxy can cool, with the onset of cooling occurring
once the gas has reached a temperature of approximately $1000 \: {\rm K}$ 
(give or take a factor of two). On the other hand, the \citeauthor{teg97} model
predicts that only those protogalaxies with $T_{\rm vir} > 1000 \: {\rm K}$ 
will cool; smaller protogalaxies, with lower virial temperatures, will simply
remain as pressure-supported gas clouds and will not form stars.

Are either of these scenarios correct? Ultimately, this depends upon whether
the approximations on which they are based are justified. This is a question 
that is best addressed through the use of more detailed numerical simulations.

\subsection{Thermal evolution: numerical simulations} 
\label{numerical} 
The biggest problem that we face when trying to simulate protogalactic 
collapse numerically is the wide range of length scales that we are required
to resolve. For example, consider the collapse of a $10^{6} \: \msun$ 
protogalaxy. This has a characteristic size (as given by the virial radius) 
of approximately $3 \: {\rm kpc}$ in comoving units, corresponding to 
$100 \: {\rm pc}$ in physical units at $z=30$. To properly simulate its
cosmological environment, we should follow the evolution of the gas and dark 
matter on scales that are two to three orders of magnitude larger (see,
for instance, the resolution study of \citealt{rgs02a}), while to resolve
star formation within it, we need to be able to follow the gas down to scales
of the order of an AU. The total dynamical range required in order to resolve
all of this within a single simulation is therefore approximately $10^{10}$,
many orders of magnitude larger than can be covered with a single fixed grid.

The simplest way in which we can obtain the required dynamical range is
to use a Lagrangian grid, i.e.\ one which moves with the fluid flow. This is 
particularly effective if one assumes that protogalaxies are spherically 
symmetric, as in this case we can use a simple one-dimensional Lagrangian 
code, such as that described by \citet{tw95}.

The earliest studies of this type were performed by \citet{pd68} and
\citet{pop75}, but the initial conditions that they adopted -- isolated,
isothermal clouds, initially in hydrostatic equilibrium -- are not 
appropriate for protogalaxies forming by dynamical collapse within an 
expanding cosmological model. More recent simulations by \citet{htl96},
\citeauthor{olv98a}~(1998ab) and \citet{sk03} begin from more appropriate 
cosmological initial conditions and all three groups obtain broadly similar
results. 

The most significant result of these simulations is the demonstration that
the dynamical evolution of a small, $\mHt$ cooled protogalaxy is {\em not} 
well described by pressure-free gravitational collapse. Instead,
pressure plays a important role, particularly for protogalaxies with masses 
near $M_{\rm J}$. It has two main effects. In the initial stages of collapse, 
when the flow is subsonic, it delays the collapse of the gas 
relative to the dark matter, resulting in a density profile which is less
centrally concentrated than would otherwise be the case. At a later time,
once the infall has become supersonic, the finite pressure leads to the
formation of an accretion shock near the virial radius of the halo.
The majority of the $\mHt$ that forms does so in the post-shock gas, and
it is the conditions there that determine whether or not the protogalaxy
is able to cool effectively \cite{htl96}. 

Of the two scenarios discussed in the previous section, the \citeauthor{teg97}
model clearly provides the better description. The major point of disagreement
is the protogalactic density profile: for simplicity, \citeauthor{teg97} 
assume a uniform density profile, while the simulation results show that the 
true profile is much closer to that of an isothermal sphere. 

Unfortunately, one-dimensional Lagrangian simulations, although simple to 
perform and quick to run, ultimately give us a rather limited view of
protogalactic formation, since they do not include many important physical
effects such as rotation and turbulence. To identify the role that these
factors play, we need to use fully three-dimensional hydrodynamical 
simulations.

The move to three dimensions necessitates a change in our computational 
strategy, as grid-based Lagrangian codes do not handle three dimensional
flows well due to the severe grid distortion that tends to occur and 
which causes a dramatic loss of accuracy. This problem can be circumvented
to some degree through the use of `hybrid' codes, which switch to
Eulerian (i.e.\ fixed) coordinates in regions of high deformation. Examples
include the codes of \citeauthor{gn95}~(1995; see also \citealt{gb96}) and 
\citet{pen98}. However, although this technique has been used to study 
protogalactic feedback (\citealt{og96}; \citealt{go97}; 
\citeauthor{rgs02a},~2002ab), it has not been used to study primordial star
formation in any detail.

Another way to avoid the grid distortion problem is to abandon the use of a
grid, and to switch to a particle-based Lagrangian technique such as 
smoothed particle hydrodynamics (SPH; see \citealt{gm77}; \citealt{lucy77}; 
\citealt{m92}). Alternatively, the required dynamical range can be obtained
using fixed grids if multiple nested grids, or some form of grid refinement
are used. Both of these approaches are discussed in more detail below.

\subsubsection{SPH simulations}
Several authors have studied the formation of protogalaxies using SPH
(\citealt{bcl99},~\citeyear{bcl02}; \citealt{fc00}; \citealt{yahs03}).
\citet{fc00} used the {\sc hydra} cosmological SPH code \cite{hydra} to study 
the collapse of uniform, spherical protogalaxies of various masses at a range
of different redshifts, in order to test the predictions of \citet{teg97}.
They obtained broadly similar results, although their values of $T_{\rm min}$
are roughly a factor of two smaller than those of \citeauthor{teg97}, a 
discrepancy which may simply be due to the different $\mHt$ cooling functions
used in the two studies.

\citeauthor{fc00} also studied protogalactic formation in a more realistic 
cosmological simulation. They demonstrated that the evolution of the most
massive object in their simulations could be divided into two main phases.
In the first phase, the halo mass is less than the Jeans mass,
pressure forces dominate, and the baryonic overdensity is small, but
non-zero. This phase corresponds to the delayed collapse phase seen in
the one-dimensional simulations discussed above.

During this phase, the mass of the halo continues to increase, driven to
a large extent by mergers with smaller dark matter halos. As the mass nears
$M_{\rm J}$, the gas density profile begins to steepen significantly as 
gravitational forces become dominant. At this stage, the gas temperature
is already significantly higher than the temperature of the IGM, while the
fractional $\mHt$ abundance in the central dense region is of order 
$10^{-4}$, two orders of magnitude larger than its initial value. 
Nevertheless, the cooling time remains longer than the dynamical timescale,
and the gas is not yet self-gravitating. 

The second phase begins once $t_{\rm cool}$ drops below $t_{\rm dyn}$ at the
centre of the protogalaxy. The central gas rapidly cools to $T \sim 150 \: 
{\rm K}$, and the consequent reduction in pressure support leads to a 
substantial increase in the central density. The gas eventually becomes 
self-gravitating at $z \simeq 20$, at which point the protogalaxy has a mass
$M \simeq 4 \times 10^{5} \: \msun$, close to the estimate of $M_{\rm min}$ at
that redshift.

Inspection of the spherically averaged temperature and density profiles of the
protogalaxy allows us to identify several distinct regions. From the outside 
in, we have: 
\begin{enumerate}
\item[(i)] Cosmological infall, terminated by an accretion shock
\item[(ii)] A broad, post-shock region, where heating from adiabatic 
compression competes with $\mHt$ cooling, and where 
$t_{\rm cool} > t_{\rm dyn}$ 
\item[(iii)] A cold, dense central core, where $t_{\rm cool} < t_{\rm dyn}$.
\end{enumerate} 
Unfortunately, the structure of the core region is not well resolved in 
\citeauthor{fc00}'s simulation, since its extent is comparable to the 
minimum smoothing length of their SPH code.

\citeauthor{bcl99}~(1999, 2002) simulated protogalactic collapse using a
modified version of the {\sc treesph} code \cite{hk89}. They concentrated on
following in detail the evolution of a single protogalaxy and consequently
adopted simplified initial conditions: a spherical overdensity, 
set into rigid rotation with a specified angular momentum and perturbed on
small scales using the Zeldovich approximation \cite{za} with a
$P(k) \propto k^{-3}$ power spectrum. By focusing on a single protogalaxy
and neglecting its cosmological environment, they were able to follow its
collapse to high densities ($n \leq 10^{8} \: {\rm cm}^{-3}$). On large
scales, their results confirm those of \citet{fc00}: the gas initially 
evolves adiabatically, is heated up to $T \sim T_{\rm vir}$ in an accretion
shock, and subsequently cools to $T \sim 100$--$200 \: {\rm K}$ in the dense 
central regions. On small scales, their greater resolution allowed them to
follow the formation of structure within the dense, cold gas. This portion of 
their results is discussed in detail in section~\ref{frag}.

Finally, \citet{yahs03} used the {\sc gadget} code \cite{syw01} to 
explore the cosmological environment in which the first protogalaxies
form. They performed the largest protogalactic simulation to date,
using 48 million SPH particles to simulate the evolution of a cosmological
volume which is $600 h^{-1}$ comoving kiloparsecs on a side. This simulation 
had a mass resolution of approximately $5000 h^{-1} \: \msun$ and a spatial 
resolution of approximately $50 h^{-1} \: {\rm pc}$, and so resolved little 
of the internal structure of the protogalaxies. On the other hand, it did 
allow a large sample of protogalaxies to be studied within a single consistent 
simulation, and was therefore a useful complement to more detailed studies of 
single protogalaxies.

\citeauthor{yahs03} found that to cool efficiently, protogalaxies in their
simulation must have masses $M \simeq 5 \times 10^{5} h^{-1} \: \msun$,
and fractional $\mHt$ abundances $f_{\mHt} \simeq 2 \times 10^{-4}$. Moreover,
while all of the protogalaxies with large $\mHt$ abundances also had large
masses, the converse was not true; some protogalaxies with masses
above $5 \times 10^{5} h^{-1} \: \msun$ did not form enough $\mHt$
to cool. Further investigation of these protogalaxies showed that they were 
gaining mass more rapidly than their cooler counterparts, leading 
\citeauthor{yahs03} to suggest that their temperatures were being kept 
high by the compressional heating associated with frequent merger activity.
If this is the case, then it implies that the ability of a particular 
protogalaxy to cool and form stars depends on its dynamical history as well
as its current mass. Further investigation of this point is clearly warranted.

\subsubsection{Multigrid simulations}
The basic idea behind a multiple grid (or multigrid) Eulerian simulation 
is to take a single 
top-level grid that is large enough to represent the whole volume of interest,
and then to supplement it with one or more levels of subgrids in regions in
which higher resolution is desired. Since much of the volume of a typical 
cosmological simulation is filled with under-dense material that is well
resolved by the top-level grid alone, this technique can dramatically improve 
the dynamical range of an Eulerian simulation for only a small increase in
its computational cost. 

In the simplest implementation of the multigrid technique, the placement of 
the grids is specified at the beginning of the simulation and does not
subsequently alter. This is the approach used in the protogalactic simulations
of \citet{aanz98a}. They used the {\sc hercules} code \cite{anc94,aanz97b} 
to study the growth of $3\sigma$ and $4\sigma$ density peaks within a large 
cosmological volume. The simulations were performed using a top-level grid 
with a resolution of $128^{3}$, together with a quarter-size subgrid with the 
same resolution, for an effective resolution of $512^{3}$. The initial 
conditions were arranged to ensure that the density peak would remain within 
the region covered by the subgrid. 

These simulations were able to resolve the basic filamentary structure of the
IGM surrounding the protogalaxies, and gave some indications that gas cooling 
within the protogalaxies was not particularly efficient. However, the 
protogalaxies remained under-resolved, rendering these conclusions uncertain.

A natural way to improve the resolution would be to add more subgrids. 
However, as one does this, it becomes increasingly difficult to ensure that
the subgrids are placed correctly, since at the beginning of the simulation
it is generally not possible to determine exactly which regions will require
very high resolution. Fortunately, this problem can be overcome by the use 
of a technique called adaptive mesh refinement. 

In an adaptive mesh simulation, the placement of subgrids is not specified
a priori. Instead, one or more refinement criteria are specified, and 
local subgrids are created as required to ensure that these criteria are 
always satisfied. Adaptive mesh codes have been used with great success in a 
number of areas of astrophysics, as discussed in the recent review of 
\citet{norm04}. Their use in the study of primordial star formation was 
pioneered by \citeauthor{abn00}~(2000, 2002) in a pair of highly influential 
papers.

In the first of these papers, \citet{abn00} used the {\sc enzo} code of 
\citeauthor{enzo1}~(1997ab) to follow the evolution of protogalactic gas
from cosmological scales down to densities of order $10^{6} \: {\rm cm}^{-3}$.
They achieved a maximum resolution of $0.4 \: {\rm pc}$ (in comoving units)
within a box that was 128 comoving kpc on a side, for a total dynamic range
of approximately $3 \times 10^{5}$. This simulation produced a protogalaxy 
with the same basic temperature profile as found in the SPH simulations: a 
cold infall region, an accretion shock at $r \sim r_{\rm vir}$, a subsequent 
broad cooling zone, and a cold, dense central region. In a subsequent 
simulation, described in \citet{abn02}, they included additional molecular 
physics (the three-body formation of $\mHt$) and followed the collapse to
significantly higher densities, eventually reaching a minimum physical 
resolution of a few tens of AU.

Although \citeauthor{abn02} present many of their results, such as the 
temperature and density profiles, in the form of spherically averaged 
quantities, they also present a number of slices through their simulations
on different scales. These demonstrate that the assumption of spherical
symmetry is a relatively crude approximation, particularly on large scales,
since much of the gas falling into the potential well of the
protogalaxy does so along a few overdense filaments, rather than in a 
spherically symmetric fashion.

The very high dynamical range achieved in their simulations also allowed
\citeauthor{abn02} to study in detail the evolution of the dense gas at the 
centre of the simulated protogalaxy. This portion of their results is 
discussed later, in section~\ref{frag}.

\subsection{Summary}
By combining the detailed results of the numerical simulations described in
the previous section with the more general theoretical considerations of 
sections~\ref{first-obj}--\ref{simple}, we are able to put together a 
reasonably comprehensive answer to the first of the questions posed in the 
introduction: when do the first protogalaxies form, and how large are they?

As we have seen, the earliest protogalaxies will form at a redshift of 
30--40, and will have masses of order $10^{4} \: \msun$. However, these
protogalaxies will form little $\mHt$ and will not be able to cool 
effectively. They are therefore extremely unlikely to form stars. The 
earliest star-forming protogalaxies will form later, at $z \sim 30$,
and will be more massive, with masses of order $10^{5}$--$10^{6} \: \msun$,
and virial temperatures of order $1000 \: {\rm K}$.  

Finally, it should be noted that these results assume a CDM-based cosmological
model. If this turns out to be an incorrect description of dark matter on 
small scales, then we should expect these numbers to change significantly. 
For instance, if some form of warm dark matter is a more appropriate 
description, then the first protogalaxies will be larger 
($M \sim 10^{7} \: \msun$) and will form at lower redshift ($z \sim 20$),
as demonstrated in the recent simulation by \citet{yshs03}. However, models 
such as this have great difficulty accounting for the high electron scattering
optical depth detected by {\sc wmap} \cite{kogut03}, and at present there 
seems little reason to prefer them over CDM.

\section{Fragmentation}
\label{frag}
Once we have established that a significant amount of protogalactic gas
can cool and condense on a cosmologically interesting timescale, the 
next step is to investigate what happens to this gas. In particular, we
would like to know whether any of it forms stars, and, if so, how
many stars form, over what timescale, and with what IMF? 

Crucial to determining this is an understanding of the degree 
to which the protogalactic gas fragments during its dynamical evolution:
does the cold gas sink into the centre of the halo, forming a single massive 
clump? Or does it break up into many smaller clumps? To address these 
questions, in the following sections I examine the effectiveness of the 
various mechanisms 
that may bring about fragmentation, and discuss the results of the most recent
numerical investigations.

\subsection{Hierarchical fragmentation}
\label{hier}
An obvious place to begin is with the force responsible for the formation 
of the protogalaxy itself: {\em gravity}. For gravitational fragmentation
to be effective, two important conditions must be met. First, any fragments 
that form must be gravitationally bound. Second, regions that begin to 
collapse due to their own self-gravity must be able to complete this collapse 
on a timescale shorter than the dynamical timescale of the flow 
in which they are embedded; otherwise, they will be disrupted before they have 
time to grow into distinct objects. 

The question of whether a particular fragment is gravitationally bound
depends, in the general case, on a number of factors: the mass of the 
fragment, its internal velocity field and pressure distribution, the 
properties of the surrounding gas flow etc.\ (see \citealt{mz92} for a 
detailed discussion). However, much of the work done on gravitational 
fragmentation in a primordial context makes the simplifying assumptions
that the only forces acting are gravity and pressure, and that the latter
can be neglected on scales larger than the Jeans length.
Although the validity of these assumptions is questionable, they provide 
a simple starting point for an investigation of protogalactic fragmentation,
so I will briefly discuss the conclusions they lead us to,
before going on to consider more detailed models.

My starting point is the work of \citet{hoyle53}. He considered the 
gravitational collapse of a homogeneous protogalaxy, and showed that on 
scales where pressure can be neglected, the second of our conditions for 
fragmentation will always be satisfied. His argument is very 
simple: gas with a density $\rho$ collapses gravitationally on a timescale 
$t_{\rm ff} \propto (G \rho)^{-1/2}$, while a perturbed region with a 
density $\rho^{\prime}$ will collapse on a timescale  
$t^{\prime}_{\rm ff} \propto \left(G \rho^{\prime}\right)^{-1/2}$. If
$\rho^{\prime} > \rho$, then $t_{\rm ff}^{\prime} < t_{\rm ff}$, and
so the perturbed region will collapse faster than the main flow.
In the protogalactic case, this 
implies that overdense regions within the protogalaxy will be able to 
collapse under their own self-gravity in less time than it takes for the 
protogalaxy itself to collapse. Therefore, the protogalaxy will fragment.

\citeauthor{hoyle53} also pointed out that one can apply precisely the same 
argument to every fragment that forms within the protogalaxy: provided that 
they contains overdense regions, and that the neglect of pressure forces 
remains appropriate, they too will fragment (as will the fragments of these 
fragments etc.). \citeauthor{hoyle53} therefore argued that the protogalactic 
gas would continue to fragment on smaller and smaller scales until pressure 
forces finally intervened to prevent further fragmentation, a scenario that 
has come to be known as {\em hierarchical fragmentation}. 

\citeauthor{hoyle53}'s original semi-quantitative argument was subsequently 
placed on a sounder mathematical basis by \citet{hunt62}, who performed a 
linear perturbation analysis of a uniform, spherical, pressure-free collapse 
and showed that any overdense perturbation would be gravitationally unstable 
and would quickly grow until it became nonlinear in less than a free-fall 
time. \citet{hunt64} expanded on this analysis by considering  second order 
terms and showed that these would accelerate the growth of overdense regions.
Similar analyses have also been made by \citet{sv62} and \citet{silk82}, with 
similar results.

\subsection{The opacity limit}
An important prediction of the hierarchical fragmentation scenario is that 
the smallest fragments will 
have sizes of the order of the Jeans length, and hence masses of the order of 
the Jeans mass, since this is the scale on which pressure balances gravity. 
However, both $\lambda_{\rm J}$ and $M_{\rm J}$ are functions of the density 
and temperature of the gas, and will change as the protogalaxy evolves. To 
identify the minimum  fragment mass, we must therefore determine how small 
$M_{\rm J}$ becomes during the collapse.

Recall that we can write $M_{\rm J}$ in terms of the gas density as
\begin{equation}
 M_{\rm J} \propto \rho^{\frac{3}{2}(\gamma_{\rm eff} - \frac{4}{3})},
\end{equation}
where the effective adiabatic index $\gamma_{\rm eff}$ is 
\begin{equation}
 \gamma_{\rm eff} = 1 + \frac{\dd{{\rm ln}\, T}}{\dd{{\rm ln}\, \rho}}.
\end{equation}
For $\gamma_{\rm eff} < 4/3$, the Jeans mass decreases with increasing 
density, while for $\gamma_{\rm eff} > 4/3$ it increases.
\citet{hoyle53} suggested that a transition from $\gamma_{\rm eff} < 4/3$ to 
$\gamma_{\rm eff} > 4/3$ would occur when the gas first became optically thick,
under the assumption that this would mark a change from isothermal evolution to
adiabatic evolution. The density and temperature of the gas at this time would
then set the minimum fragment mass. This basic idea -- that it is the opacity 
of the gas which sets a lower limit on the mass of a fragment and therefore 
on the mass of a star -- has become known as 
{\em opacity-limited fragmentation}, and has been 
studied by a number of authors.

\citet{llb76} and \citet{silk77b} both follow \citeauthor{hoyle53} in 
assuming that the minimum mass is reached at the moment that a fragment first 
becomes optically thick. They also assume that the fragment is in thermal 
balance at this time, with compressional heating balanced by radiative cooling.
These assumptions provide them with two equations relating $T_{\rm F}$, 
$\rho_{\rm F}$ and $\kappa_{\rm F}$ (the temperature, density and opacity of
the fragment at the moment that it becomes optically thick):
\begin{eqnarray}
 \kappa_{\rm F} \rho_{\rm F} \frac{\lambda_{\rm J}(\rho_{\rm F}, 
 T_{\rm F})}{2} & = & 1, \label{opthick} \\
 \Gamma_{\rm c}(\rho_{\rm F}, T_{\rm F}) & = & \Lambda_{\rm r}(\rho_{\rm F}, 
 T_{\rm F}, \kappa_{\rm F}),
\end{eqnarray} 
where $\Gamma_{\rm c}$ is the compressional heating rate and $\Lambda_{\rm r}$ 
is the radiative cooling rate. 

We can use these equations to express the minimum fragment mass $M_{\rm F}$ 
in terms of a single unknown -- for instance, \citet{llb76} write it in terms 
of the opacity as
\begin{equation}
 M_{\rm F} = 2.5 \times 10^{-3} \mu^{-16/7} \left( 
\frac{\kappa_{0}}{\kappa_{\rm F}} \right)^{1/7} \: \msun,
\end{equation}  
where $\mu$ is the mean molecular weight and $\kappa_{0}$ is the opacity due 
to Thomson scattering -- but to fully 
determine $M_{\rm F}$, we need an additional piece of information. In Hoyle's
original analysis, this comes from the assumption that the radiative cooling
is dominated by Lyman-$\alpha$ emission, which fixes the temperature at 
approximately $10^{4} \: {\rm K}$. On the other hand, \citet{silk77b} 
considers a case where the cooling and opacity are both dominated by dust, 
in which case the observed dust temperature provides the additional 
information required. In general, however, we can only determine $M_{\rm F}$ 
if we know something of the previous thermal history of the gas.
 
\citet{rees76} studied the opacity limit from a different viewpoint, arguing 
that the essential requirement for continued isothermal evolution is that a
fragment be able to radiate away its gravitational binding energy in less than
a free-fall time, and that opacity is only important inasmuch as it limits the
maximum radiative rate, which cannot exceed that of a black body of a similar
temperature. This fact can be used to derive the minimum temperature that a 
fragment must have in order to radiate sufficient energy, which in turn can be 
used to determine $M_{\rm F}$:
\begin{equation}
 M_{\rm F} = M_{\rm c} \, \mu^{-9/4} f^{-1/2} \left( 
\frac{kT_{\rm F}}{m_{\rm p} c^{2}}  \right)^{1/4}, \label{rees_mf}
\end{equation}
where $M_{\rm c}$ is the Chandrasekhar mass, $m_{\rm p}$ is the mass of a 
proton and $f$ is a radiative efficiency factor, defined as the ratio of the
actual radiation rate to the black-body rate:
\begin{equation}
 f = \frac{\int F_{\nu} \: \dd{\nu}}{\int \pi B_{\nu} \: \dd{\nu}}.
\end{equation}
From Equation~(\ref{rees_mf}), we see that if $f \sim 1$, then $M_{\rm F}$ 
will be of the order of a solar mass (or less, if $T_{\rm F}$ is very small), 
while if  $f \ll 1$, as may occur if the cooling is dominated by a few narrow 
emission lines, then $M_{\rm F}$ may be of the order of tens or hundreds of 
solar masses.

More recently, \citet{mi99} have reconsidered the conditions under which 
isothermal evolution comes to an end. They show that this will inevitably 
occur once the radiative cooling rate becomes unable to keep pace with the 
compressional heating rate, and that this may take place in either the 
optically thin or optically thick  regime, depending on the details of the 
collapse, but is unlikely to coincide with
the instant at which $\tau = 1$. This suggests that a better procedure 
for determining $M_{\rm F}$ is to follow the actual thermal evolution of the 
gas.

\subsection{Simple numerical models}
Various authors have attempted to calculate $M_{\rm F}$ by modelling the 
thermal evolution of the collapsing protogalactic gas. One of the earliest was
\citet{yon72}, who constructed an evolutionary track for the gas in 
density-temperature space by assuming that it always satisfied the following 
conditions:
\begin{eqnarray}
t_{\rm cool} & = &  t_{\rm ff}, \\
t_{\mHt}     & = & {\rm max}(t_{\rm ff}, t_{\rm rec}),
\end{eqnarray}
where $t_{\mHt}$ is the $\mHt$ formation timescale, given by
\begin{equation}
t_{\mHt} = \frac{n_{\mHt}}{k_{\Hm} \, n_{\rm e} \, n_{\rm H}},
\end{equation}
where $k_{\Hm}$ is the rate coefficient for the formation of $\Hm$ by 
radiative association of electrons and \hi (reaction~\ref{h2f1}),
and $t_{\rm rec}$ is the recombination timescale, given by
\begin{equation}
t_{\rm rec} = \frac{1}{k_{\rm rec} \, n_{\Hp}},
\end{equation}
where $k_{\rm rec}$ is the rate coefficient for radiative recombination.
\citeauthor{yon72} used this technique to calculate the evolution of 
$M_{\rm J}$ until either the gas became optically thick or became hot enough 
to collisionally dissociate $\mHt$. $M_{\rm F}$ could then be computed simply 
by finding the minimum value of $M_{\rm J}$ reached along the evolutionary
trajectory. \citeauthor{yon72} found that in small protogalaxies, the minimum
value was reached shortly before the gas became hot enough to collisionally
dissociate $\mHt$ and that $M_{\rm F} \simeq 60 \: \msun$; in larger 
protogalaxies, the greater column density of $\mHt$ caused the gas to become 
optically thick before collisional dissociation could occur, and the 
resulting value of $M_{\rm F}$ was considerably larger.

A more common approach is to specify the form of $\rho(t)$ in advance, often by
constructing some extremely simplified dynamical model for the protogalaxy, and
then to use this as an input into a more detailed chemical and thermal model. 
Many such models exist 
\cite{hutch76,silk77a,carl81,hys81,pss83,ls84,lahav86,vb87,dao89,sun96}; 
I will discuss only a few notable examples.

\citet{hutch76} studied the thermal evolution of a variety of spherical and 
spheroidal protogalaxies using a very simplified chemical model consisting of
only four reactions -- formation of $\mHt$ via the $\Hm$ pathway 
(reactions~\ref{h2f1}--\ref{h2f2}), plus radiative recombination of hydrogen
\setcounter{saveqn}{\value{equation}}
\setcounter{equation}{\value{reaction}}
\renewcommand{\theequation}{\mbox{R\arabic{equation}}}
\begin{equation} 
 \Hp + \me \rightarrow \mH + \gamma, 
\end{equation} 
and collisional dissociation of $\mHt$ by $\mH$:
\begin{equation} 
 \mHt + \mH \rightarrow 3 \mH. \label{coll-dissoc}
\end{equation} 
\setcounter{reaction}{\value{equation}}
\setcounter{equation}{\value{saveqn}}
\renewcommand{\theequation}{\arabic{equation}}
His cooling function was equally simple, and included only Compton cooling 
and $\mHt$ rotational cooling. The calculations were terminated once $\mHt$ 
dissociated. \citeauthor{hutch76} found a minimum fragment mass of 
approximately $200 \: \msun$, within a factor of a few of 
\citeauthor{yon72}'s result.

\citet{silk77a} and \citet{carl81} both investigated the effects of including
Lyman-$\alpha$ cooling in the model, and showed that it has a dramatic effect. 
The reason is that the collisional dissociation of $\mHt$ no longer results 
in a permanent transition to adiabatic evolution. Instead, the gas heats up 
adiabatically for a short while, until its temperature reaches 
$10^{4} \: {\rm K}$, following which it again
begins to evolve isothermally. This second phase of isothermal evolution is
eventually terminated once the fragments become optically thick in the 
continuum. It allows $M_{\rm F}$ to reach much smaller values than would
otherwise be possible. 
\citet{silk77a} estimated the minimum fragment mass to be 
approximately $0.3 \: \msun$; \citet{carl81}, using a more detailed 
treatment, found $M_{\rm F} \sim 0.5 \: \msun$. 

Finally, \citet{pss83} investigated the importance of three-body $\mHt$ 
formation (reactions~\ref{h23a}--\ref{h23b}) and showed that at high densities 
($n > 10^{8} \: {\rm cm}^{-3}$), these reactions become very effective, 
rapidly converting the
bulk of the hydrogen to molecular form. This dramatically increases the $\mHt$
cooling rate, delaying the collisional dissociation of $\mHt$ until much higher
densities are reached, and consequently lowers $M_{\rm F}$. For the 
representative 
example of a $5 \times 10^{4} \: \msun$ cloud, 
\citet{pss83} found a minimum fragment mass of $M_{\rm F} \sim 0.1 \: \msun$.

The main lesson to draw from these attempts is the importance of an accurate 
treatment of the thermal evolution of the gas, which requires a 
comprehensive treatment of the microphysics -- comparison of the results of 
\citet{hutch76} with those of \citet{silk77a} or \citet{pss83} demonstrates
the danger of using an oversimplified chemical or thermal model. However,
the specific predictions of these models rely upon the accuracy of the 
assumptions made regarding the density evolution, and we have already seen 
that simple collapse models generally do not perform well in this respect.
More importantly, these predictions rely on 
the correctness of the basic assumptions underlying the hierarchical 
fragmentation scenario. Unfortunately, there are good reasons to believe
that these assumptions are not correct, as I discuss in the next section.

\subsection{The case against hierarchical fragmentation}
\label{case_agin}
Recall that the hierarchical fragmentation scenario is based on two major
assumptions: first, that the balance between gravity and pressure is the
only determinant of whether a fragment is gravitationally bound, and second,
that the gas flow on scales larger than the Jeans length can be approximated
by pure gravitational free-fall. The first of these assumptions implies that
all fragments with $M > M_{\rm J}$ will be gravitationally bound; the second,
that the gas can quickly fragment down to the Jeans scale, provided that it
starts from moderately uniform initial conditions. Clearly, both of these
assumptions represent substantial simplifications. However, the real 
question is whether the simplified picture remains accurate; in other words, 
did hierarchical fragmentation actually occur in real protogalaxies?

One important argument against this scenario was first advanced by 
\citet{lay63}. He argued that as the protogalactic gas fragmented, the
individual fragments would tend to acquire angular momentum from the
gravitational torques exerted on them by their neighbours. If this angular
momentum was subsequently conserved during the evolution of the fragment,
then it would limit the extent to which it could contract, and would help
to stabilize it against further fragmentation.

A convenient way to parameterize this is in terms of the dimensionless spin
parameter
\begin{equation}
 \lambda \equiv \frac{J |E|^{1/2}}{GM^{5/2}} \simeq 
 \frac{v_{\rm rot}}{v_{\rm ff}},
\end{equation}
where $J$ and $E$ represent the fragment's total angular momentum and 
total energy respectively. If angular momentum is conserved, then
the fragment's rotational velocity will scale as $v_{\rm rot} \propto R^{-1}$ 
(where $R$ is the size of the fragment), and since its free-fall velocity 
will scale as $v_{\rm ff} \propto R^{-1/2}$, this implies that 
$\lambda \propto R^{-1/2}$; in other words, the fragment will spin faster as
it contracts. The fragment will become centrifugally supported once
$\lambda = 1$, and so the contraction of the fragment will stop once
it reaches a size
\begin{equation}
 R = \lambda_{0}^{2} R_{0} \label{spin}
\end{equation} 
where $\lambda_{0}$ is the initial spin parameter of the fragment, and
$R_{0}$ is its initial size.

\citet{lay63} estimated the initial spin parameter to be of order unity, and
therefore argued that any fragments that formed would be barely distinct from
the background gas, and would inevitably collide and coalesce before the end
of the protogalactic collapse. \citet{hunt64}, however, disagreed and argued
that the mutual gravitational torques between the fragments would not affect
their spins. He noted that in a freely-falling, isothermal, inviscid gas, 
Kelvin's circulation theorem would apply, and so therefore the vorticity of
a fragment would change only due to its expansion or contraction; the 
gravitational field would have no direct effect. In this analysis, 
Equation~(\ref{spin}) still applies, but $\lambda_{0}$ depends purely on the
initial vorticity of the gas forming the fragment, and is substantially 
less than one.

Regardless of the correct value of $\lambda_{0}$, it is clear from this 
analysis that at some point the fragments will become centrifugally 
supported, provided that they conserve angular momentum. Once this occurs,
the assumption of free-fall collapse made by Hoyle, Hunter and many others
is clearly no longer appropriate. Moreover, centrifugally supported 
fragments will tend to flatten into rotating disks, which have 
different stability properties from those of collapsing spheres, as 
discussed at some length in \citet{lar85}. Specifically, one can define
a parameter
\begin{equation}
Q = \frac{c_{\rm s} \kappa}{\pi G \mu}
\end{equation}
where $\kappa$ is the epicyclic frequency of the disk and $\mu$ is its surface
density, such that disks with $Q > Q_{\rm crit}$ are stable against 
gravitational fragmentation\footnote{Note that this $Q$ is analogous to the
\citet{toom64} stability parameter for a stellar disk.}. The value of 
$Q_{\rm crit}$ depends to some extent on the equation of state of the gas, 
but typically $Q_{\rm crit} \sim 0.5$--0.6 \cite{glb65,lar85}. Furthermore,
even if the first generation of centrifugally supported fragments remain 
unstable (i.e.\ if they have $Q < Q_{\rm crit}$), subsequent generations 
will be more highly stabilized (\citeauthor{lar84}~1984,~1985), provided that
the evolution is isothermal. Therefore, rather than the successive fragmentation
envisioned in the hierarchical fragmentation scenario, one may instead that 
find only one or two generations of fragments form, with rotation quickly 
acting to suppress fragmentation on smaller scales.

Another serious critique of the hierarchical fragmentation scenario was
put forward by \citet{toh80}. He pointed out that there is a fundamental
difference between the growth of perturbations in a pressure-free collapse,
such as that considered by \citet{hunt62}, and in a pressurized collapse. In
a pressure-free collapse, overdensities are gravitationally unstable on 
{\em all} scales, and begin to grow immediately. In a pressurized collapse,
on the other hand, only overdensities with masses larger than the initial
Jeans mass can grow to begin with. Although smaller overdensities will 
subsequently become unstable and begin to grow as $M_{\rm J}$ decreases 
during the protogalactic collapse, the onset of their growth is delayed.
This effect is particularly pronounced if the initial Jeans mass is close to
the mass of the protogalaxy, as in this case small perturbations will be
unable to grow until after the protogalaxy has already collapsed by a 
significant amount. 

Based on this, \citeauthor{toh80} argued that one of the fundamental 
assumptions of the hierarchical fragmentation scenario -- the rapid 
fragmentation of the gas down to scales of order of the Jeans mass -- 
is not correct, since $M_{\rm J}$ may vary faster than the gas can 
respond. In particular, he argued that the minimum mass of a fragment at
the time that the gas becomes optically thick will not be equal to the Jeans
mass at that moment, since overdensities with $M \sim M_{\rm J}$ will only 
just have become unstable, and will not yet have had the chance to grow. 
Instead, the minimum fragment mass will correspond to the Jeans mass at some
earlier time, and will therefore be significantly larger than is predicted by
the models discussed in the previous section.

Another way to consider this issue is to examine the dispersion relation
arising from the perturbation analysis. For the classical Jeans
analysis of plane wave density perturbations in an infinite uniform medium,
one obtains
\begin{equation}
 \omega^{2} = c_{\rm s}^{2} k^{2} - 4 \pi G \rho, \label{disp_rel}
\end{equation}
and the same relation holds for spherically symmetric density perturbations
of the form $r^{-1} {\rm sin} \: kr$ \cite{lar85}. In the pressure-free case,
$c_{\rm s} = 0$, and the growth rate of perturbations is scale-free. On the
other hand, in the pressurized case, $c_{\rm s}$ is non-zero and 
the growth rate increases with decreasing wavenumber, reaching a maximum
for $k = 0$. In other words, large-scale perturbations grow faster than 
small-scale perturbations, suggesting that for the latter to win out and
to cause
fragmentation to occur, they must start with substantially higher densities.
Although Equation~(\ref{disp_rel}) applies directly only to a rather idealized
protogalaxy, it is reasonable to expect to find similar behaviour in the more
general case.

A final problem with the hierarchical fragmentation scenario is that it
assumes that the gas is initially uniform (which implies that $t_{\rm ff}$ is 
the same everywhere), and remains so during the collapse. However, this is
unlikely to be the case. Even if the gas begins its collapse from a uniform
state, it will tend to become centrally concentrated during the course of the
collapse \cite{bs68,lar73,toh82} as a pressure gradient builds up between the 
centre and the edge of the protogalaxy. This is significant, because a 
centrally concentrated gas cloud is far more stable than a uniform cloud
with regard to the growth of small self-gravitating perturbations, as a 
number of authors have demonstrated \cite{arny66,ms80,ss88,lc89}.
Although many of these analyses assume a high 
degree of symmetry, which makes unclear the extent to which the results will
apply in more realistic models of collapse, they do provide another indication
that fragmentation in real protogalaxies will be far less effective than 
the hierarchical fragmentation scenario assumes.

\subsection{Other forms of fragmentation}
In view of the doubts raised in the previous section concerning the 
effectiveness of fragmentation driven purely by gravity, it is worthwhile
to spend some time examining two other processes which may bring about
fragmentation of the protogalactic gas: thermal instability and supersonic
turbulence.

\subsubsection{Thermal instability}
\label{ti} 
So far we have considered the gas pressure purely as a stabilizing force,
resisting the action of gravity. However, if the protogalactic gas is 
thermally unstable, then the pressure can itself drive fragmentation within 
the gas. 

Thermal instability occurs when small perturbations to the density and/or
temperature of a region of gas cause the subsequent temperature evolution 
of that region to differ significantly from that of the unperturbed gas.
The case of interest to us is the one in which the perturbations cause 
accelerated cooling of the perturbed region. This was first investigated 
by \citet{pa53}, who derived an instability criterion
\begin{equation}
 \left(\frac{\partial{\cal L}}{\partial T} \right)_{\rho} < 0,
\end{equation}
where ${\cal L} = (\Lambda - \Gamma)/\rho$ is the specific heat loss function,
under the assumption that the perturbations were isochoric (i.e. that the
gas maintained a constant density). \citet{field65} pointed 
out that in this case, pressure gradients would develop in the gas, and 
that thermal instability could therefore drive dynamical flows. He also 
argued that on small scales, the gas would rapidly respond to any change 
in temperature by changing its density, so as to keep its pressure constant.
In other words, small perturbations would evolve isobarically rather than
isochorically. \citet{field65} derived an instability criterion for the 
isobaric case
\begin{equation}
  \left(\frac{\partial{\cal L}}{\partial T} \right)_{p} = 
  \left(\frac{\partial{\cal L}}{\partial T} \right)_{\rho} 
  - \frac{\rho_{0}}{T_{0}} \left(\frac{\partial{\cal L}}{\partial \rho} 
  \right)_{T} < 0, 
\end{equation}
where $\rho_{0}$ and $T_{0}$ are the unperturbed density and temperature.
This is the correct criterion for perturbations with wavelengths 
$\lambda_{\rm F} < \lambda < \lambda_{\rm c}$, where 
$\lambda_{\rm c} = c_{\rm s} t_{\rm cool}$ is the distance travelled by a
sound wave in a single cooling time in the unperturbed medium, and where
$\lambda_{\rm F}$ is the Field length, given by
\begin{equation}
 \lambda_{\rm F} = \left(\frac{\kappa T}{\Lambda}\right)^{1/2},
\end{equation}
where $\kappa$ is the coefficient of thermal conduction. For perturbations
with $\lambda > \lambda_{\rm c}$, Parker's isochoric criterion applies,
while for perturbations with  $\lambda < \lambda_{\rm F}$, thermal conduction 
completely suppresses the instability.

If we apply these criteria to primordial gas, while keeping the chemical
composition of the gas fixed, then we find that the gas is always thermally
stable. However, if we allow the chemical composition of the gas to vary as
we perturb the temperature and density, then various {\em chemo-thermal} 
instabilities become possible. The simplest of these occurs in very hot gas
with $10^{5} < T < 10^{7} \: {\rm K}$. Within this temperature regime, cooling 
via \heii line emission increases sharply with decreasing temperature as 
helium recombines from \heiii to \heii, leading to a pronounced thermal 
instability. This instability has been investigated in the context of galaxy 
formation by 
\citeauthor{ml90}~(\citeyear{ml90},~\citeyear{ml96}; see also 
\citeauthor{lm92},~\citeyear{lm92},~\citeyear{lm00})
but seems unlikely to play a significant role in the evolution of the first 
protogalaxies, as the protogalactic gas never becomes hot enough to trigger 
it. 

At lower temperatures, various instabilities associated with the formation 
and dissociation of $\mHt$ may occur. The first of these was identified by
\citet{sy77} and analyzed in more detail by \citet{ys79}. It occurs in gas
with a temperature in the range $2000 \simless T \simless 4000 \: {\rm K}$
(with a slight dependence on density) and is caused by the collisional 
dissociation of $\mHt$. Within this temperature range, the equilibrium 
abundance of $\mHt$ is a sensitive function of temperature, due to the 
strong temperature dependence of the collisional dissociation rate 
(reaction~\ref{coll-dissoc}). Therefore, gas which has a slightly lower 
temperature than its surroundings will have a higher $\mHt$ abundance, and 
hence a higher cooling rate. If this leads to a further drop in temperature 
and increase in $\mHt$ abundance, then an instability results. Gas cooler 
than about $2000 \: {\rm K}$ does not suffer from the instability because 
the collisional dissociation rate becomes too small to significantly affect 
the $\mHt$ abundance, which therefore loses its strong temperature dependence,
while in gas hotter than about $4000 \: {\rm K}$, the $\mHt$ abundance becomes
too small to provide effective cooling.

\citet{silk83} identified a related instability that appears once the 
three-body formation of $\mHt$ becomes effective and occurs for a similar 
reason: a small decrease in the temperature and the associated increase in
the density lead to an enhanced $\mHt$ abundance and higher cooling rate, 
which further perturb the temperature. This instability vanishes if the gas
becomes fully molecular or becomes optically thick to $\mHt$ line emission.

Finally, another potential instability has recently been identified by 
\citet{ra04}. This one occurs at very high densities ($10^{14} < n < 10^{15} 
\: {\rm cm}^{-3}$) and is due to a combination of the onset of 
collisionally-induced emission from the $\mHt$ (which is highly 
sensitive to the gas density and which quickly comes to dominate the 
$\mHt$ cooling rate), and a renewed phase of collisional dissociation 
within the dense gas (which at slightly lower densities is fully 
molecular). However, the resulting instability is very sensitive to the 
temperature of the gas, as can be seen clearly from figures~7~and~8 of 
\citet{ra04}, and it remains to be seen whether this instability will 
actually occur in practice.

Since thermal instability is capable of creating dense structure in the gas 
on all scales larger than the Field length, we might expect it to profoundly 
influence the ability of the gas to fragment. However, in practice, it is 
unlikely to be of great importance in primordial gas. There are two main
reasons for this. First, the $\mHt$-related instabilities discussed above
all operate within a fairly restricted range of temperatures. This 
significantly limits the size of the temperature contrasts that can be 
created, which in turn limits the size of the resulting density contrasts, 
which can therefore be disrupted more easily by other processes such as 
turbulence \cite{abn02}. Second, thermal instabilities grow on the cooling 
timescale, 
$t_{\rm cool}$, and will therefore cause significantly restructuring of the
gas only when $t_{\rm cool} \ll t_{\rm dyn}$. However, \citet{oy03} and 
\citet{ra04} demonstrate that thermally unstable protogalactic gas typically 
has $t_{\rm cool} \simeq t_{\rm dyn}$, implying that the instabilities do
not have sufficient time to grow. This conclusion is supported by the results 
of \citet{abn02}, whose simulation includes gas in the thermally unstable 
regime associated with three-body $\mHt$ formation, but who find no 
indication that this instability has any significant dynamical effect.

\subsubsection{Supersonic turbulence} 
Another way to create dense structures in the gas without the assistance of
gravity is by compressing it in shocks. In particular, if the velocity field
of the gas is turbulent and supersonic, then large density enhancements
can be created as the gas is repeatedly shocked. This is a process that has
received considerable attention in a Galactic context, since there is 
substantial observational evidence for the existence of supersonic 
turbulence in interstellar gas on all scales larger than a few tenths of
a parsec. Rather than attempting to summarize all of this material here -- 
a hopeless task -- I refer the reader to the recent comprehensive reviews of 
\citet{mk04}, \citet{es04} and \citet{se04}, and restrict my discussion to a 
few points of particular relevance to primordial star formation.

Simulations of supersonic turbulence in self-gravitating, isothermal gas 
have been performed by a number of groups 
\cite{ogs99,khm00,hmk01,glso03,bbb03,li04}. The gas in these simulations
rapidly develops a highly inhomogeneous structure, with a density 
probability distribution function (PDF) which is approximately log-normal.
Dense, gravitationally bound cores form in the highest density regions, 
with a mass spectrum that also appears to be log-normal and which resembles
the stellar IMF. The efficiency with which cores form depends upon the 
properties of the turbulence, and is lower in models with more power on 
smaller spatial scales \cite{khm00}, but it appears to be very difficult to 
completely suppress fragmentation: this would require strong turbulence
on very small scales, which would rapidly decay away \cite{sog98,ml99} unless 
driven by some form of energy input on those scales.

A series of attempts have been made to relate the stellar IMF directly to
the statistical properties of interstellar turbulence 
\cite{lar81,fle82,elm93,pad95,pnj97,my00,pn02}, which
would allow the IMF to be predicted in environments such as early 
protogalaxies for which no direct observational determinations exist. However,
none of these attempts have met with widespread acceptance, and research in 
this area is still ongoing.

Extension of these results to primordial gas is further complicated by the
fact that most studies of turbulent fragmentation assume an isothermal 
equation of state. This is a reasonable approximation for gas in local 
molecular clouds, since its cooling time is very short, but it is
unlikely to be appropriate for primordial gas with $t_{\rm cool} \sim 
t_{\rm dyn}$. Since there are indications that relatively small changes
in the equation of state can have a large effect on both the shape of the
density PDF \cite{ps98} and on the numbers and masses of self-gravitating 
cores that form \cite{lkm03}, a straightforward extrapolation from the
isothermal results appears unwise. Work in this area is ongoing.

\subsection{Numerical simulations}
As the previous sections hopefully make clear, a number of different factors
influence the ability of the protogalactic gas to fragment. While we can gain
considerable insight into the physics of the individual processes through
the use of simple analytical models, for a proper understanding of how the 
various different processes interact within a real protogalaxy we are 
currently forced to turn to numerical simulations.

\subsubsection{Simulations of local star formation}
Before discussing the results of simulations designed specifically to study
protogalactic fragmentation and primordial star formation, it seems worthwhile
to examine what we can learn from simulations designed primarily to study
local star formation. 

One important thing that we have learnt from this kind of simulation is the 
vital importance, when studying gravitational collapse and 
fragmentation, of resolving the Jeans length throughout the simulation. This 
was convincingly demonstrated by \citet{true97}, who showed that if this 
criterion is not met, then completely artificial fragmentation of the gas 
can result. This implies that simulations that fail to meet this criterion 
cannot be used to make meaningful predictions about gravitationally-driven 
fragmentation. Although \citet{true97} restricted their attention to 
grid-based simulations, it has been shown that SPH simulations suffer from 
a very similar problem \cite{bb97,whit98}.

Also of interest are the results of a set of simulations performed by
\citeauthor{ti99a}~(1999ab). They used high resolution SPH simulations to 
study the isothermal collapse of a set of uniform spherical clouds with 
varying ratios of thermal to gravitational energy, parameterized by
\begin{equation}
 \alpha_{0} = \frac{5c_{\rm s}^{2} R_{0}}{2 G M},
\end{equation}
where $R_{0}$ is the initial radius of the cloud, and of rotational to 
gravitational energy, parameterized by
\begin{equation}
 \beta_{0} = \frac{\Omega_{0}^{2} R_{0}^{3}}{3 G M},
\end{equation}
where $\Omega_{0}$ is the initial angular velocity. Previous analytical 
and numerical work \cite{toh81,mhn84} suggested that such a cloud would
collapse to a disk, with a flatness that depended on the product 
$\alpha_{0}\beta_{0}$, and that for $\alpha_{0}\beta_{0} < 0.12$, this disk
would subsequently fragment. However, this criterion is clearly not correct 
when $\beta_{0}$ is very small, as it predicts that a cloud with 
$\alpha_{0} > 1$ should collapse and fragment, whereas in reality such a 
cloud would have a mass smaller than the Jeans mass, and would not collapse.
Moreover, the analytical derivation of this criterion assumes that the 
collapse is homologous, while in reality collapsing clouds would tend to 
become centrally concentrated.

\citeauthor{ti99a} find three possible outcomes for their simulated clouds. 
When $\alpha_{0}$ and $\beta_{0}$ are both small, the cloud collapses to
a thin disk and fragments, much as was previously envisaged. As $\alpha_{0}$ 
increases, however, the thickness of the disk also increases, and once its
flatness -- defined simply as the ratio of its radius to its scale height --  
falls below a value of approximately $4\pi$, the disk no longer fragments.
Further increases in $\alpha_{0}$ lead to a final state that increasingly
comes to resemble the so-called Larson-Penston similarity 
solution.~\footnote{The Larson-Penston solution is an asymptotic similarity 
solution for the isothermal collapse of a sphere,
independently derived by \citet{lar69} and \citet{pen69}, which describes 
the collapse at late times and/or at small distances from the centre, 
when the influence of the boundary conditions has become negligible. It can 
be derived numerically from the governing equations of the flow if one 
assumes that the flow is smooth (i.e.\ that there are no shocks) and that 
the central velocity is zero; this derivation can also be generalized to the 
case of a polytropic equation of state $P = K \rho^{\gamma}$ \cite{lar69}.
Although other similarity solutions are possible \cite{shu77,hunt77,ws85}, 
the Larson-Penston solution appears to provide the best fit to the results
of numerical simulations of isothermal spherical collapse \cite{fc93}.}
Finally, for sufficiently large $\alpha_{0}$, collapse is entirely suppressed.

In the low $\beta_{0}$ limit, the boundary between fragmentation and 
self-similar collapse occurs for $\alpha_{0} \simeq 0.5$; with increasing
$\beta_{0}$, the boundary moves to smaller $\alpha_{0}$, and is given 
approximately by $\alpha_{0} = 0.55 - \beta_{0}$ for $0 < \beta_{0} < 0.3$.

Although the initial conditions for these simulations are highly idealized, 
they do provide strong support for some of the criticisms of the hierarchical
fragmentation scenario discussed in section~\ref{hier}, and demonstrate that
gravitational fragmentation does not appear to be as effective as originally
anticipated, particularly in clouds with large values of $\alpha_{0}$. 

\citet{ti01} extended this analysis, in a more limited fashion, to the 
case of protogalactic collapse. They performed three large SPH simulations 
of the collapse of uniform, spherical protogalaxies, all with the same initial
temperature ($T_{0} = 150 \: {\rm K}$) and rotation parameter 
($\beta_{0} = 0.25$), but with differing masses: $M = 10^{6}$, 
$3 \times 10^{6}$, $10^{7} \: \msun$, implying that
$\alpha_{0} = 0.18, 0.09$ and 0.04 respectively.
Rather than assuming that the gas remains
isothermal, \citeauthor{ti01} follow its chemical and thermal evolution
during the collapse. Based on the previous isothermal results, one would 
expect fragmentation to occur in all three simulations, but in fact 
fragmentation does not occur in the $10^{6} \: \msun$ protogalaxy, which 
instead simply forms a single dense central core. This difference from the
isothermal case is presumably due to the fact that cooling is less efficient in
these more realistic model protogalaxies, which therefore evolve as if they 
had larger values of $\alpha_{0}$. If this interpretation is correct, then
it suggests that we can treat \citeauthor{ti99a}'s isothermal fragmentation
criterion as a necessary, but not sufficient, criterion for fragmentation
within primordial gas. As we shall see below, this interpretation is 
consistent with the results of more detailed protogalactic simulations.

\subsubsection{Filamentary collapse}
The propensity of gravitationally collapsing spheres of gas to settle into
disks even when $\beta_{0}$ is small, noted by \citeauthor{ti99a} and by many
other authors, is not unexpected, since we have known for a long time that
small departures from spherical symmetry are steadily magnified during
free-fall collapse \cite{lms65}. Moreover, flattened, disk-like clouds will
generally fragment into filamentary structures (\citeauthor{mnh87a}~1987ab)
which only subsequently fragment into clumps. In light of this, a number of 
authors have considered the problem of filamentary collapse in primordial 
gas. 

A pressure-supported filament (i.e.\ one which is not collapsing or 
expanding radially) is gravitationally unstable to perturbations along the 
axis of the filament. Moreover, it is possible to show that for an isothermal 
filament, the fastest growing perturbation is the one with wavelength 
$\lambda_{\rm c} \sim \pi R$, where $R$ is the scale radius of the filament,
defined as:
\begin{equation}
 R = \left( \frac{M}{\pi \rho_{0}} \right)^{1/2}, \label{scale_r}
\end{equation}
and where $M$ is the mass per unit length and $\rho_{0}$ is the central 
density \cite{stod63}. A similar result can be derived for a polytropic
filament \cite{lar85}. However, filaments formed during protogalactic collapse
are unlikely to start in hydrostatic equilbrium in the radial direction.
\citet{u96} argue that in that case, it is necessary to follow the radial 
evolution of the filament until such time as the contraction timescale, given 
by $t_{\rm con} = \rho_{0} / \dot{\rho_{0}}$, exceeds the fragmentation 
timescale, $t_{\rm frag} \sim (G\rho_{0})^{-1/2}$; at this point, the 
equilibrium analysis can be applied to give an estimate of the resulting 
fragment mass.

\citet{u96} used this approach to study the fragmentation of primordial 
filaments with a range of values of $M$. They adopted an initial temperature,
density and molecular fraction appropriate to gas that had already undergone
significant cooling and collapse within a protogalaxy, consistent with their
assumption that the filaments had formed in a fragmenting disk. They followed 
the subsequent chemical and thermal evolution of the filament in some detail, 
but treated the density evolution in an approximate fashion: the scale radius 
was assumed to evolve as
\begin{equation}
 \ddot{R} = - \frac{2G}{R}\left[ M - M_{\rm c}(T) \right],
\end{equation}
where $M_{\rm c}$ is the mass per unit length of an equilibrium isothermal
cylinder \cite{os64}
\begin{equation}
 M_{\rm c}(T) = \frac{2kT}{\mu m_{\rm H} G}, 
\end{equation}
in which case the density then follows from Equation~(\ref{scale_r}). 
They studied a number of different collapses, with values of $M$ ranging from
1--$2 \: M_{\rm c}$. They found that as $M$ increased, there was a 
corresponding increase in the density at which the filament fragmented, and
a consequent decrease in the fragment mass $M_{\rm F}$, which fell from
$100 \: \msun$ for $M = M_{\rm c}$ down to $2 \: \msun$ for 
$M = 2 M_{\rm c}$.

\citet{nu99} reconsidered this problem and improved on the \citet{u96} analysis
in several important respects. Most significantly, they replaced the 
approximate treatment of the density evolution used by \citeauthor{u96} with
a more accurate treatment based on a one-dimensional hydrodynamic simulation
of the collapse of the filament. They also made use of a more extensive 
chemical model and considered a wider range of initial conditions. They 
found that the collapse led to one of two possible outcomes, depending on the
initial temperature and the value of $f \equiv M_{\rm c}/M$. In the majority
of the models, $\mHt$ cooling was effective at the start of the simulation
and remained so until the gas became optically thick at high density 
($n \sim 10^{11} \: {\rm cm}^{-3}$). This allowed the filaments to collapse 
dynamically to high density, fragmenting only once cooling became ineffective,
and producing fragments with masses $M_{\rm F} \sim 1$--$10 \: \msun$.
However, in models where the initial temperature was low 
($T_{0} = 100 \: {\rm K}$), $\mHt$ cooling was not immediately effective, and 
the initial evolution of the filament was adiabatic. In models where the mass 
per unit length was large ($f > 2$), collapse could persist until the 
temperature increased to a point at which $\mHt$ cooling became effective, 
following which the evolution of the filament would continue much as if it
had started with a higher initial temperature. If the initial mass per unit
length were small, however, then collapse would very quickly come to an end,
resulting in an equilibrium filament with a relatively low central density
($n \sim 10^{5} \: {\rm cm}^{-3}$) that would produce much larger fragments 
with masses of a few hundred $\msun$.

\citet{nu01} further improved the treatment of this problem by performing
two-dimensional axisymmetric hydrodynamical simulations of filamentary
collapse, which allowed them to follow the fragmentation numerically, rather
than estimating its effects analytically. They again examined a wide range of
initial conditions, although in light of their previous results, they 
restricted their attention to filaments with initial temperatures 
$T_{0} \geq 300 \: {\rm K}$.
Once more, they found two possible outcomes. In filaments with a low initial 
density ($n \simless 10^{5} \: {\rm cm}^{-3}$), fragmentation occurred prior
to the onset of three-body $\mHt$ formation and the 
resulting fragments were large, with masses $M_{\rm F} \sim 100 \: \msun$.
On the other hand, in filaments with a high initial density, fragmentation 
is delayed until after the gas has become optically thick, resulting in 
much smaller fragments of mass $M_{\rm F} \sim 1$--$2 \: \msun$.

The potential role played by $\HD$ molecules in filamentary collapse has been 
investigated by \citet{ui00} and \citet{nu02}. \citeauthor{ui00} assumed
that the filaments formed in a shock-bounded sheet, and therefore adopted 
initial conditions appropriate to primordial gas which has cooled rapidly from 
temperatures $T \gg 10^{4} \: {\rm K}$. As previously demonstrated by 
\citet{ms86} and \citet{sk87}, hydrogen recombination lags behind cooling 
in such gas, resulting in an elevated fractional ionization that allows a 
substantial $\mHt$ fraction ($f_{\mHt} \sim 10^{-2}$) to build up. 
\citeauthor{ui00} demonstrate that in these conditions, the fractional 
abundance of $\HD$ would be of order $10^{-5}$, and 
showed that this amount of $\HD$ is enough to cool the filament to 
$50 \: {\rm K}$ and to keep it evolving isothermally at this temperature 
until the $\HD$ lines become optically thick. They argued that this allows 
very low mass fragments to form, with masses $M_{\rm F} \sim 0.01$--$0.1 \: 
\msun$.

\citet{nu02} examined collapse from a wider range of initial conditions
than \citet{ui00}, and showed that $\HD$ cooling would only be significant 
if the initial 
$\mHt$ and $\HD$ abundances were both large. However, even in this case, they
obtained a much larger fragment mass than \citet{ui00}, finding
$M_{\rm F} \sim 10 \: \msun$. They ascribe this difference in part to their
more detailed treatment of optical depth effects, and in part to the fact that
\citeauthor{ui00} assumed that the minimum fragment mass would equal the Jeans
mass, rather than the mass contained within the fastest growing perturbation, 
which in this case is about an order of magnitude larger. 

Filamentary collapse has also been studied by \citet{fl02} and \citet{fp03},
who examined the effects of including a magnetic field. \citet{fl02} 
used a dynamical treatment similar to that of \citet{u96} to show that even a 
relatively weak axial magnetic field would soon provide enough pressure to
halt the collapse, resulting in the formation of massive fragments, with 
sizes ranging from $M_{\rm F} \sim 60 \: \msun$ for an initial field strength
of $10^{-9} \: {\rm G}$ up to $M_{\rm F} \sim 6000 \: \msun$ for 
an initial field strength of $10^{-7} \: {\rm G}$. However, \citet{fp03} 
subsequently showed that if the effects of ambipolar diffusion were also 
included, then the field would be far less effective at slowing the 
collapse, since the fractional ionization of the gas in the filament is
too low to keep the field strongly tied to the gas.

Ultimately, in spite of the attention paid to filamentary collapse, the 
relevance of these results to fragmentation in real protogalaxies remains 
unclear. The main concerns are that all of these simulations assume initial 
conditions that are far more smooth and symmetrical than will actually be
the case in a real collapse, and that they neglect a number of potentially
important effects such as rotation and turbulence. 

\subsubsection{Three-dimensional simulations of protogalactic collapse}
Relatively few 3D simulations of protogalaxy formation have been
performed to date, and of these the only ones with sufficient dynamical range
to study fragmentation within the newly formed protogalaxy are the SPH 
simulations of \citeauthor{bcl99}~(1999,~2002), and the adaptive mesh 
refinement simulations of \citeauthor{abn00}~(2000,~2002), both of which 
were discussed previously in section~\ref{numerical}.

As previously noted, \citeauthor{bcl99} study collapse from somewhat idealized
initial conditions: a single, isolated overdensity, in rigid rotation with
spin parameter $\lambda$, and set to collapse at some specified redshift 
$z_{\rm c}$. They begin their simulations at $z = 100$, and evolve from then
until shortly after the protogalaxy has virialized. By focusing on a single
protogalaxy in this way they are able to achieve a high mass resolution. The
precise resolution depends on the mass of the protogalaxy simulated and the 
number of SPH particles used, but for their fiducial case of a 
$2 \times 10^{6} \: \msun$ protogalaxy with a baryon fraction of 5\%, 
simulated with $16384$ particles, \citeauthor{bcl99} achieve a mass resolution
of approximately $200 \: \msun$. To avoid the numerical difficulties that would
otherwise force the simulations to be halted once the Jeans mass fell below
this value \cite{bb97,whit98}, \citeauthor{bcl99} make use of a sink particle 
technique \cite{bbp95}. SPH particles with densities greater than a threshold
value $n_{\rm th} = 10^{8} \: {\rm cm}^{-3}$ and which are in regions of 
converging flow ($\nabla \cdot {\bf v} < 0$) are removed from the simulation,
and replaced with one or more sink particles. One sink particle is created for
each individual collapsing region, with a mass and momentum equal to the sum
of the masses and momenta of the particles that it has replaced. Once created,
sink particles continue to interact with the surrounding gas particles, and 
can accrete them if they lie within two smoothing lengths and satisfy the
criteria above. Further details of the algorithm are given in \citet{bcl02}.

\citet{bcl99} present results from a single simulation of protogalactic 
collapse. The protogalaxy they simulate has a total mass of 
$2 \times 10^{6} \: \msun$, a baryon fraction of 5\%, a spin parameter
$\lambda = 0.05$, and collapses at a redshift $z = 30$. Following the initial
sequence of compression, shock and subsequent cooling and settling that has
already been described in section~\ref{numerical}, \citeauthor{bcl99} find
that the cooled gas settles into a flattened central disk, with a radius
of approximately $15 \: {\rm pc}$ and thickness of $2 \: {\rm pc}$. This disk
rapidly breaks up into about a dozen dense clumps, with masses ranging from
200--$10^{4} \: \msun$. The gas in the disk has a mean temperature of 
approximately $200 \: {\rm K}$ (although there is considerable scatter about 
this value), and a density of order $10^{4} \: {\rm cm}^{-3}$. The gas in the 
clumps is somewhat hotter ($T \sim 500 \: {\rm K}$) and substantially denser,
with densities ranging all the way up to $n_{\rm th}$. \citet{bcl99} find no
evidence for further fragmentation of the clumps, but since they are soon 
replaced in the simulation by sink particles, they are unable to rule it out.
However, further fragmentation of the clumps appears unlikely: they have 
relatively large ratios of thermal to gravitational energy ($\alpha_{0} 
\simeq 0.5$ for a typical clump), and so based on the \citeauthor{ti99a} 
criterion, one would not expect them to fragment. 

\citet{bcl02} discuss the results of a more extensive range of simulations, 
which sample a wider range of initial conditions. They find that the 
thermodynamic behaviour of the gas is very similar in each of their 
simulations. In every case, the gas cools rapidly until it reaches a 
temperature and density at which cooling becomes ineffective. In a gas 
dominated by $\mHt$ cooling, this occurs at $T \sim 200 \: {\rm K}$ and 
$n \simeq 10^{4} \: {\rm cm}^{-3}$: below this temperature, the $\mHt$ cooling
rate falls off exponentially, while above this density, collisional 
de-excitation of the excited levels of $\mHt$ significantly reduces its 
effectiveness as a coolant. Since fragmentation typically occurs after the 
gas has reached this state, the fragment masses tend to lie close to the
Jeans mass corresponding to this temperature and density.

In contrast, the morphology of the cool gas is far more sensitive to the
initial conditions of the simulation. In most cases, it is disk-like,
but the size and visual appearance of the disks alter significantly as the 
spin parameter and the collapse redshift are varied. A notable exception is
the single low-mass protogalaxy simulated by \citet{bcl02}, which had a total
mass $M = 2 \times 10^{5} \: \msun$, and the usual baryonic fraction of 5\%.
This had a larger degree of pressure support than the more massive 
protogalaxies, and settled into a spheroidal, quasi-hydrostatic equilibrium
configuration at a density of order $10^{2} \: {\rm cm}^{-3}$. 
\citeauthor{bcl02} found no evidence for fragmentation within this protogalaxy,
although a single massive dense clump did form in its centre, much as in 
the high $\alpha_{0}$ simulations of \citet{ti99a}. 

Finally, \citeauthor{bcl02} also examined the effects of $\HD$ cooling and
showed that it made very little difference to the outcome of the simulation.
This appears to be due to the fact that although $\HD$ becomes the dominant
coolant in low temperature gas, it never becomes an {\em effective} coolant 
-- the cooling time of the gas remains significantly longer than the 
dynamical time, and so the gas does not become significantly cooler than it 
would if the $\HD$ were not included in the simulation.

\citeauthor{abn00}~(2000,~2002) pursed a rather different strategy in their
study of protogalactic collapse. Rather than simulating collapse from a range
of different initial conditions, they instead focused on simulating a single
example in great detail, starting from realistic initial conditions within
a large simulation volume, and following the collapse to higher densities than
those reached in \citeauthor{bcl99}'s SPH simulations. On large scales, their
results agree with those of other simulations of protogalactic collapse, as
I have already discussed in section~\ref{numerical}. On smaller scales,
they find an accumulation of cold gas within the central ten parsecs of the
protogalaxy, much as \citeauthor{bcl99} do. However, the morphology of this
gas is not at all disk-like -- it is more like a slightly flattened spheroid,
although less symmetric than this description suggests. \citeauthor{abn00} 
find no evidence for any fragmentation of this cool gas, beyond the formation
of a single dense core of mass $M \sim 100 \: \msun$ at its centre.

At the moment that it forms, this central core has a similar temperature and
density to the surrounding cool gas, namely $T \simeq 200 \: {\rm K}$ and
$n \simeq 10^{4} \: {\rm cm}^{-3}$, giving it a ratio of thermal to  
gravitational energy $\alpha_{0} \simeq 0.5$. It is also rotating slowly, 
with $\beta_{0} = 0.01$. These values suggest that even if the core were to
collapse isothermally, it would be unlikely to fragment further. In fact,
the collapse of the core is not isothermal: instead, the gas rapidly heats
up to a temperature of about $800 \: {\rm K}$. Indeed, it is this sharp rise
in temperature, as much as anything, that distinguishes the core from its
surroundings, as its density profile merges smoothly with the surrounding gas.

This rise in temperature, which is also seen in the cores that form in the 
\citeauthor{bcl02} simulations, albeit to a lesser degree, is a natural 
consequence of the reduced efficiency of $\mHt$ cooling at high densities.
Above a critical density of approximately $10^{4} \: {\rm cm}^{-3}$, the
$\mHt$ cooling rate begins to scale as $\Lambda_{\mHt} \propto n$, while
the compressional heating rate scales as $\Gamma_{\rm comp} \propto n^{3/2}$,
and so the latter eventually begins to dominate, causing the core temperature
to increase. 

In their original simulation, \citeauthor{abn00} followed the evolution of 
the core only up to a density of $10^{8} \: {\rm cm}^{-3}$; as their chemical 
model did not include 3-body $\mHt$ formation, which becomes effective at this
density, any results from higher densities would have been highly inaccurate.
In their subsequent simulation, they included the three body reactions,
allowing them to follow the collapse of the core to much higher densities.
They were eventually forced to stop at a density of 
$10^{13} \: {\rm cm}^{-3}$, because the $\mHt$
rotational and vibrational lines were becoming optically thick, and their
assumption of optically thin cooling was therefore no longer valid. 

\citeauthor{abn02} found no evidence for fragmentation of the central core in
either of their simulations. In particular, the thermal instability associated
with $\mHt$ formation and discussed in section~\ref{ti} appears to have only 
a minor effect on the evolution of the core, and does not cause it to 
fragment. Turbulence is also ineffective at driving fragmentation, since the 
collapse is predominantly subsonic, only becoming marginally supersonic 
within the central 10--20$\: {\rm AU}$ near the end of the simulation.

Finally, \citeauthor{abn02} show that the core never becomes rotationally
supported, and that its final rotational velocity is about half of the 
Keplerian orbital velocity $v_{\rm Kep} = (GM / r)^{1/2}$.
This is one of their most surprising
results, as it implies that angular momentum is being transferred outwards
during the evolution of the core. Indeed, \citeauthor{abn02} are able to 
demonstrate directly that this is occurring (see figure~4a of \citealt{abn02}).
This result inevitably raises the suspicion that it is due to some purely 
numerical effect, such as numerical shear viscosity \cite{nwb80}. However, 
\citeauthor{abn02} find that the details of the angular momentum transfer
are independent of the hydrodynamic algorithm used and of the spatial 
resolution (provided that the simulation continues to satisfy the 
Truelove criterion). This suggests that the effect that they find is real,
but of course is not yet conclusive; it would be extremely useful to be 
able to reproduce this result with another hydrodynamical code, ideally
one based on a fundamentally different algorithm, such as SPH.

\citet{abn02} ascribe the angular momentum transfer to the action of shocks
during the collapse, but this conclusion is open to question since, as noted 
above, the infall is subsonic in most of the core. Another possibility is
that angular momentum is transferred by tidal interactions with external
mass concentrations \cite{lar02}. Ultimately, to develop an understanding of 
the physics underlying this effect, we are likely to require additional high 
resolution adaptive mesh simulations \cite{norm03}.

\subsubsection{The optically thick phase}
To follow the evolution of the gas beyond the density reached by \citet{abn02},
it is necessary to solve a radiative transfer problem for the optically 
thick $\mHt$ line emission, in order to be able to calculate the correct 
cooling rate. Unfortunately, an exact solution of this problem within a 
three-dimensional hydrodynamical simulation is not currently feasible, since 
it is essentially a seven-dimensional problem (three spatial dimensions, two 
angles plus frequency and time). Approximate methods, such as the {\sc otvet} 
formalism of \citet{gnab01} will help in the near future, but so far 
the only simulations that have been performed of this last stage of 
protogalactic evolution have been forced to assume spherical symmetry, purely
for reasons of efficiency. This unfortunately renders them mute on such topics
of interest as whether the efficient outward transfer of angular momentum found
by \citeauthor{abn02} continues at higher densities, or whether the core 
fragments into a close binary or multiple system rather than a single star.

The first detailed simulations of the evolution of the core in the optically
thick phase were performed by \citet{on98}. They used an explicit 
one-dimensional Lagrangian hydrodynamical code to simulate the collapse of a
small number of model cores with different masses and densities. At gas 
densities below $10^{15} \: {\rm cm}^{-3}$, they followed the chemical 
evolution of the gas using a simplified chemical model based on that of 
\citet{pss83};
at higher densities, chemical equilibrium was assumed, and the chemical 
abundances were obtained from solution of the coupled Saha equations. Cooling
from both $\mHt$ line emission and collision-induced continuum emission was 
included; the latter dominates at very high densities. \citeauthor{on98} 
computed the radiative transfer of this emission using the tangent ray method 
\cite{hr71}, under the assumption that line transfer and continuum 
transfer can be decoupled. To further simplify the calculation, they assumed
that the diffusion approximation holds in regions that are highly opaque in 
the continuum, and that line cooling from this gas was negligible.

\citeauthor{on98} found that after a short initial transient, the evolution 
of each of their model cores was essentially the same, and so they presented 
detailed results for only a single example: a polytropic core of mass 
$M = 100 \: \msun$
and density at the half-mass radius $n_{\rm h} = 10^{6} \: {\rm cm}^{-3}$.
As it collapsed, this core quickly developed a self-similar density profile
with $\rho \propto r^{-2.2}$, and the collapse as a whole was well described
by a Larson-Penston type similarity solution for a gas with an equation of
state $p = K \rho^{\gamma}$, where $K = 4.2 \times 10^{11}$ (in cgs units),
and $\gamma = 1.09$. The gas in the centre of the collapsing core soon
became optically thick in the $\mHt$ lines, but this did not immediately 
lead to the evolution becoming adiabatic, as enough cooling was possible 
through the optically thin wings of the lines, as well as in the continuum via
collision-induced emission, to maintain $\gamma_{\rm eff} < 4/3$ for an 
extended period. Eventually, however, the core temperature became high
enough to thoroughly dissociate $\mHt$, and the evolution became fully
adiabatic. This occurred for a central density $n_{\rm c} = 10^{22} \: 
{\rm cm}^{-3}$, and lead to the formation of a small hydrostatic core,
with mass $M = 5 \times 10^{-3} \: \msun$, at the centre of the flow. This
core rapidly became fully ionized, and was bounded by an accretion shock at 
a radius of 2 AU, and it seems natural to identify it as a protostar.
Unfortunately, \citeauthor{on98} were unable to follow its subsequent 
evolution, as the Courant timestep became prohibitively small once the core 
had formed, forcing them to terminate their simulation.

More recently, \citet{ripa02} also simulated the optically thick collapse
phase. Their basic approach was similar to that of \citet{on98}, but with two
major improvements: they included a term in the momentum equations 
corresponding to the force exerted on the gas by the scattered $\mHt$ 
emission, and they used a more detailed model for the chemical evolution of the
gas and the behaviour of the equation of state at very high densities, based
on \citet{scvh95}. In addition, they also examined a wider range of initial
conditions. Despite this, they found essentially the same behaviour as
\citeauthor{on98}. The evolution of the model cores was strongly convergent
and soon became well described by a Larson-Penston type similarity solution.
This self-similarity lasted until a small hydrostatic core of mass 
$3 \times 10^{-3} \: \msun$ formed at the centre of the flow.

\subsection{Summary}
In  the introduction, I posed a number of questions concerning the evolution 
of gas within newly formed protogalaxies, namely: does the gas fragment? If 
so, how large are the fragments? And when and why does fragmentation stop?

Much of the work that has been done on primordial star formation assumes some
version of the hierarchical fragmentation scenario, in which fragmentation
is highly efficient and is terminated only by the transition of the gas from
isothermal to adiabatic evolution. In these models, the main uncertainties 
are the cause of this transition -- chemical changes or fragment opacity? --
and the temperature and density at which it occurs.

However, as I outlined in section~\ref{case_agin}, there are a number of 
reasons to believe that hierarchical fragmentation does not occur in real 
protogalaxies as various effects combine to inhibit fragmentation. This
conclusion is supported by the results of the simulations of 
\citeauthor{ti99a}~(1999ab,~2001), \citeauthor{bcl99}~(1999,~2002) and
\citeauthor{abn00}~(2000,~2002): in each of these simulations there is
at most a single episode of fragmentation, and no evidence for any 
subfragmentation (i.e. fragmentation of the fragments). Moreover, in some
of these simulations, such as \citeauthor{abn00}~(2000,~2002), the use of
the word `fragmentation' to describe the evolution of the gas is misleading:
the single `fragment' that forms is actually just the central dense core
of a more extended density distribution.

Why is it that fragmentation is so inefficient? Inefficient fragmentation
appears to be a natural outcome of quasi-spherical collapse in gas with a 
high ratio of thermal to gravitational energy. During such a collapse, the
large thermal pressure will create a strongly peaked density distribution,
even if the gas is initially quite uniform. It will also suppress small-scale 
collapse until the density of the gas has increased by a large factor, since
the local Jeans mass scales as $M_{\rm J} \propto n^{-1/2}$ in isothermal
collapse. In combination, these effects imply that the gas at the centre
of the protogalaxy will quickly come to have both the smallest Jeans mass 
and the shortest free-fall time and therefore, unless its collapse is 
delayed or halted in some way, it will continue to collapse all the way
to protostellar densities before the bulk of the gas has had the opportunity 
to fragment.
If this interpretation is correct, it suggests that further fragmentation
may occur within, say, the \citeauthor{abn00} simulations, if they were
continued past the point at which the first star forms. However, since this
first star will exert a strong feedback on its surroundings on a short 
timescale \cite{on99,gb01,byh03,wan03} it appears unlikely that any further 
fragmentation would in fact occur.

The efficacy of fragmentation could be enhanced by delaying the collapse
of the densest gas, giving the lower density gas more time in which to
fragment. As \citeauthor{bcl02} demonstrate, rotation can do this to some 
extent, but the outward transfer of angular momentum identified by 
\citeauthor{abn02} makes it
less effective than simple estimates would suggest. Strong perturbations
arising from thermal instability or supersonic turbulence could also boost 
fragmentation, but in practice neither process is particularly effective
in primordial gas.

The few fragments which do form typically have initial masses of a few
hundred $\msun$. This particular mass scale appears to be a consequence
of the thermodynamics of the gas. At densities less than the $\mHt$ critical
density of $10^{4} \: {\rm cm}^{-3}$, $\mHt$ cooling is efficient, and the 
gas can cool to a minimum temperature of about 200~K. At higher densities,
$\mHt$ cooling becomes less efficient, and the gas heats up. These values
of density and temperature therefore mark the point at which isothermal
evolution comes to an end and $\gamma_{\rm eff}$ first exceeds one, and
so it is not surprising that the minimum fragment mass corresponds 
approximately to the value of the Jeans mass at this density and temperature.

The major uncertainty that remains concerns the behaviour of the collapsing
gas in the optically thick regime. It may continue to collapse 
quasi-spherically, in which case we would expect a single, low-mass 
protostellar core to eventually form, as in the simulations of \citet{on98} 
and \citet{ripa02}. On the other hand, it may form a gravitationally unstable
disk, in which case fragmentation into a binary or multiple system would 
appear to be more likely. Resolution of this uncertainty awaits the 
development of an accurate and efficient way of treating the thermal 
evolution of the optically thick gas.

\section{Protostellar accretion and the final stellar mass}
\label{accr}
The results of the simulations of \citet{on98} and \citet{ripa02} suggest
that the initial mass of a primordial protostar may be very small, no more
than a few thousandths of a solar mass. However, this small initial protostar
will be surrounded by a large envelope of infalling gas, some fraction of 
which will inevitably be accreted by the protostar.

If mass loss from the protostar is negligible (which seems to be a good 
approximation even for very massive metal-free stars -- see 
\citealt{mck03}), then the final stellar mass $M_{*}$ is related to the 
initial protostellar mass $M_{\rm pr}$ by
\begin{equation}
 M_{*}(t) = M_{\rm pr} + \int_{0}^{t} \dot{M}(t´) \dd{t´}.
\end{equation}
The final mass is therefore determined by the evolution of the mass 
accretion rate $\dot{M}(t)$ over the lifetime of the star. This in turn is 
influenced both by the properties of the gas surrounding the star -- the 
amount of gas available, its temperature and angular momentum, etc.\ -- and by 
the effects of feedback  from the star onto the gas, in the form of radiation 
and outflows. 

\subsection{Accretion in the absence of feedback}
Protostellar feedback is complicated to model, and so it is easier to begin 
by considering models of protostellar accretion that do not include its 
effects. Since feedback will act to reduce the accretion rate, and hence also 
the final stellar mass, these models allow us to place an upper limit on the 
ultimate mass scale of the first stars.

One possible approach to determining the accretion rate is to construct a
simplified model for the collapsing protostellar core from which an 
approximation to the true accretion rate can be derived. For instance, if we
assume that the protostellar core is isothermal and spherically 
symmetric, then there exists an entire family of similarity solutions that 
could potentially be used to describe the collapse \cite{hunt77,ws85}, of 
which the most familiar are the Larson-Penston solution \cite{lar69,pen69} 
and the Shu solution \cite{shu77}. 

\begin{figure}
\centering
\epsfig{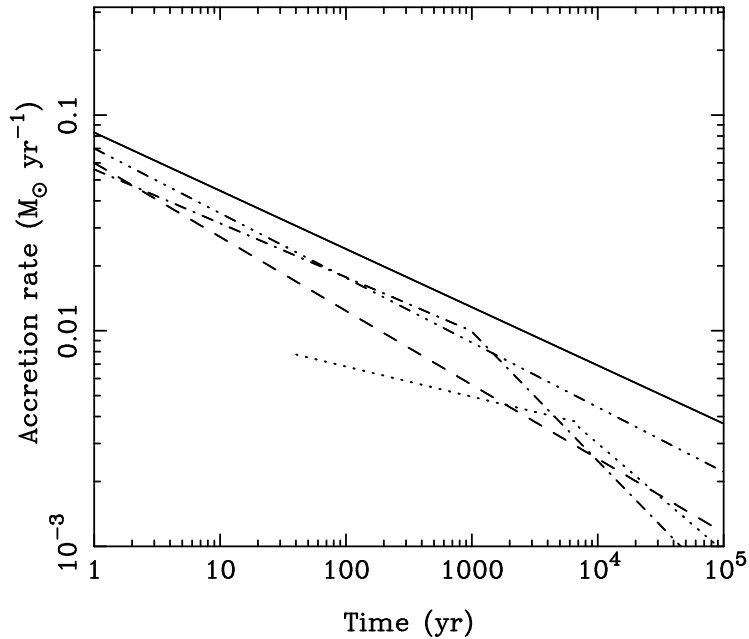}
\caption{The time-dependent accretion rates predicted by various models of
protostellar accretion. Solid line -- Omukai and Nishi (1998); dashed line
-- Ripamonti {\em et al.}~(2002); dotted line -- Abel, Bryan, and Norman 
(2002); dash-dotted line -- Bromm and Loeb (2004); dash-dot-dot-dotted 
line -- Tan and McKee (2004). In plotting the Tan and McKee rate, I have 
assumed $K^{\prime} = 1$. For $t \simless 40 \: {\rm yr}$, the predicted 
accretion rate of Abel, Bryan, and Norman (2002) depends on the behaviour 
of gas on scales close to or below the resolution limit of their simulation, 
and is therefore highly uncertain.}
\label{accr_rate}
\end{figure}

This approach has recently been applied to primordial star formation by
\citet{tm04}. They model the accretion flow as a spherical, isentropic
polytrope, and derive an accretion rate that is a function of three
parameters: the entropy parameter $K = p / \rho^{\gamma}$, the polytropic
index $\gamma_{p}$ (which, for an isentropic flow, is equal to the adiabatic 
index $\gamma$), and $\phi_{*}$, a numeric parameter of order unity, which 
is related to the initial conditions of the flow. \citeauthor{tm04} normalize
these parameters based on the numerical results discussed in the previous
sections, and set $\gamma_{p} = 1.1$, $\phi_{*} = 1.43$ and 
$K = 1.88 \times 10^{12} K^{\prime}$ (in cgs units), where 
\begin{equation}
 K^{\prime} = \left(\frac{T_{\rm eff}}{300 \: {\rm K}}\right) 
 \left( \frac{n_{\rm H}}{10^{4} \: {\rm cm}^{-3}} \right)^{-0.1}, 
 \label{kprime}
\end{equation}
and where the effective temperature $T_{\rm eff} = P_{\rm eff} / (nk)$ 
includes the small contribution to the effective pressure made by subsonic
turbulence. With these values, the accretion rate 
becomes~\footnote{Strictly speaking, the accretion rate derived by 
\citet{tm04} is for accretion onto both the protostar and its associated
accretion disk. Moreover, they also allow for the possibility that 
protostellar feedback may reduce the amount of gas reaching the centre of
the system. However, for ease of comparison between their results and those of
the other authors discussed in this section, I have neglected these 
complications for the time being.}
\begin{equation}
 \dot{M} = 7.0 \times 10^{-2} {K^{\prime}\,}^{3/2} 
 \left( \frac{t}{\rm 1 \: yr} \right)^{-0.30} \: \msun \: {\rm yr}^{-1}.
\end{equation}
This is plotted in figure~\ref{accr_rate} for the case of $K^{\prime} = 1$. 

Another obvious approach is to simulate the accretion flow numerically, 
but, as previously discussed, an accurate 3D simulation remains impractical
due to the expense of the radiative transfer calculations.
We are therefore forced to choose between 
simulating the radiative transfer and the cooling accurately, at the cost of 
restricting the hydrodynamics to one dimension, or simulating the 
hydrodynamics correctly, at the cost of oversimplifying (or simply 
neglecting) the radiative transfer effects.

 Two important examples of the first approach are the simulations of 
\citet{on98} and \citet{ripa02}, discussed at the end of section~\ref{frag}. 
To recapitulate: \citeauthor{on98} use a spherically-symmetric Lagrangian 
code to simulate protostellar core formation within initially polytropic 
clouds, and include a thorough treatment of radiative transfer within the 
$\mHt$ lines and in the continuum, using the tangent ray method \cite{hr71}.
\citeauthor{on98} find that prior to core 
formation the flow is well described by a Larson-Penston type similarity 
solution; specifically, the solution corresponding to 
$K = 4.2 \times 10^{11} \: ({\rm cgs})$ and $\gamma = 1.09$. 
They are unable to continue their simulations after the formation of the 
protostellar core, as the Courant timestep in the central regions becomes 
prohibitively small. However, if the same Larson-Penston type solution 
were to apply after core formation, then the resulting accretion rate 
would be
\begin{equation}
 \dot{M} = 8.3 \times 10^{-2} \left( \frac{t}{\rm 1 \: yr} \right)^{-0.27} 
\: \msun \: {\rm yr}^{-1}
\end{equation}
and the stellar mass would grow as
\begin{equation}
 M_{*} = 0.11 \left( \frac{t}{\rm 1 yr} \right)^{0.73} \: \msun.
\end{equation}

The simulations of \citet{ripa02} are similar in design to those of 
\citet{on98}, but incorporate several significant improvements. Specifically, 
\citeauthor{ripa02} include:
\begin{enumerate}
\item[(i)] A term in the momentum equation corresponding to the radiative 
force per unit mass 
\begin{equation}
 f_{\rm rad} = \frac{1}{c} \int \kappa_{\nu} F_{\nu} \dd{\nu},
\end{equation}
where $\kappa_{\nu}$ and $F_{\nu}$ are the opacity and specific energy flux 
at frequency $\nu$.
\item[(ii)]  An improved treatment of chemistry and thermodynamics at very 
high densities ($n > 10^{21} \: {\rm cm^{-3}}$), based on \citet{scvh95}, that 
accounts for non-ideal effects such as pressure ionization.
\item[(iii)] A `frozen core' approximation, which keeps the central mass 
shells fixed in space once their infall velocities fall below a specified
value ($v/v_{\rm ff} < 10^{-3}$) and their temperatures exceed 
$5 \times 10^{4} \: {\rm K}$. This approximation allows the simulations 
to avoid the 
worst of the Courant timestep limitations, and hence enables them to be 
continued into the period after core formation.
\end{enumerate}

Prior to core formation, there is good agreement between the results
of \citet{on98} and \citet{ripa02}, confirming that the flow at this initial 
stage is well described by a Larson-Penston type 
similarity solution. Differences appear, however, once the protostellar 
core has formed. For initial conditions comparable to those studied in 
\citet{on98}, \citeauthor{ripa02} obtain an accretion rate that is 
approximately
\begin{equation}
 \dot{M} = 6.0 \times 10^{-2} \left( \frac{t}{\rm 1 yr} \right)^{-0.343} 
 \: \msun \: {\rm yr}^{-1},
\end{equation}
which is smaller than the \citeauthor{on98} rate and falls off more rapidly. 

\citeauthor{ripa02} also find that the accretion rate is sensitive to the 
initial conditions of the simulation: clouds with higher initial temperatures 
produce cores with 
larger accretion rates, even though the simulations strongly converge at late 
times. The difference in rates is large enough to produce a difference of a 
factor of a few in $M_{*}$ by the end of the simulations, which follow only 
the first 10 years after core formation. At later times, we would expect the 
difference to be even more pronounced.

Turning to the `hydrodynamical' approach, the best examples are the simulation
by \citet{abn02}, discussed in detail in the previous section, and the
recent work by \citet{bl04}. In both cases, the full hydrodynamical problem is
solved, using adaptive mesh refinement in the former case, and SPH with 
particle resampling  \cite{kw02,bl03} in the latter. Additionally, chemistry 
and cooling are followed accurately up to the point at which opacity effects 
begin to dominate. At this point, the two treatments diverge. 
\citeauthor{abn02} halt their simulation once the optical depth at line 
centre of the main $\mHt$ cooling lines exceeds 10, at which point the 
maximum gas density is approximately $10^{13} \: {\rm cm}^{-3}$, and the 
size of the dense core is a few tens of AU. They estimate the subsequent 
accretion rate based on a calculation of the accretion timescale, 
$t_{\rm acc}$, at the end of their simulation. They take $t_{\rm acc}$ to be
\begin{equation}
 t_{\rm acc} = \frac{M(r)}{\dot{M}(r)} = \frac{M(r)}{4\pi \rho(r) r^{2} 
|v_{r}(r)|}, 
\end{equation}
where $v_{r}$ is the radial velocity of the gas, and they assume that the
stellar mass at time $t$ is simply the mass of gas with $t_{\rm acc} \leq t$.
The resulting inferred accretion rate is plotted in figure~\ref{accr_rate}.

\citet{bl04} are also unable to follow the flow to very high densities, since
they too neglect opacity effects when calculating their $\mHt$ cooling rate.
However, unlike \citet{abn02}, they are not forced to halt their simulation
once the gas becomes optically thick. Instead, they replace the SPH particles
representing the dense, optically thick gas with sink particles of the type 
described earlier. Sink particle creation is handled much as in \citet{bcl02};
the main difference is that the density threshold for sink creation is much 
higher, being set at $n_{\rm th} = 10^{12} \: {\rm cm}^{-3}$. Note that while
the code is capable of creating multiple sink particles, in practice only a 
single sink is required. By using a sink particle, they sacrifice the ability
to follow the further evolution of the high density gas, and the ultimate 
formation of the protostar, but in return can continue to study the gas flow
on larger scales for an extended period. On the assumption that all of the
gas that is accreted by the sink particle will in reality be accreted by the
protostar, they derive a protostellar accretion rate that is approximated by
a broken power-law:
\begin{equation}
 \dot{M} = \left\{ \begin{array}{lr}
 5.6 \times 10^{-2} \left(\frac{t}{1 {\rm yr}} \right)^{-0.25} & 
 \hspace{.5in} t \leq 10^{3} \: {\rm yr} \\
 & \\
 6.3 \times 10^{-1} \left(\frac{t}{1 {\rm yr}} \right)^{-0.6} & 
 \hspace{.5in} t > 10^{3} \: {\rm yr}
                   \end{array} \right. 
\end{equation} 

Although the two hydrodynamical models agree well at late times (as can be
seen by a comparison of their predicted accretion rates, which are plotted
in figure~\ref{accr_rate}), at early times there is considerable 
disagreement, with \citet{bl04} predicting a much higher initial 
accretion rate than \citet{abn02}. Unfortunately, with only a single 
example of each simulation available, it is not possible to say how much of 
this disagreement can be attributed to the difference in simulation methods, 
and how much simply reflects natural variation between the accretion rates in
different protogalaxies. Further simulations along these lines would clearly
be valuable. 

An important open question, which has yet to be studied numerically, is what
role angular momentum plays in the final stages of accretion onto the 
protostar. In particular, we would like to know whether angular momentum 
continues to be transported efficiently outwards, as it is in the simulations
of \citet{abn02}, or whether it remains approximately constant once the
accretion flow becomes supersonic, as assumed by \citet{tm04}. This is 
important because in the former case the bulk of the accretion will occur 
directly onto the surface of the protostar, while in the latter case 
accretion will occur primarily through a circumstellar accretion disk.

By it very nature, this problem is not one that can be tackled using 
one-dimensional simulations; a full three-dimensional treatment is called for.
However, disk formation, if it occurs, will take place after the flow has
become optically thick, since the initial disk radius is predicted to be only
a few AU \cite{tm04}. Absent a sudden increase in computing power
sufficient to allow us to treat the coupled radiative transfer and 
hydrodynamics accurately using an algorithm such as that outlined in
\citet{hn03}, the best approach is probably to look for some
approximate treatment that succeeds in capturing the essential behaviour of 
the flow, even if this turns out to be somewhat inaccurate. One possible
approximation is outlined in 
\citet{ra04}. They derive an $\mHt$ cooling rate for optically thick gas by
considering the simplified problem of radiation escaping from a spherically
symmetric, collapsing protostellar core. The resulting cooling rate is well
fit by
\begin{equation}
 L_{\mHt, {\rm thick}}(T) =  L_{\mHt, {\rm thin}}(T) \, {\rm min}
 \left(1, (n/n_{0})^{-\beta} \right),
\end{equation}
where $n_{0} = 8 \times 10^{9} \: {\rm cm}^{-3}$, $\beta = 0.45$, and where
$ L_{\mHt, {\rm thin}}(T)$ is the $\mHt$ cooling rate in optically thin gas.
\citeauthor{ra04} demonstrate that this simple approximation performs well in 
comparison to the full radiative transfer solution used in \citet{ripa02};
however, its accuracy in the three dimensional case is currently unknown.
More work along these lines is clearly called for if we are to make progress
on solving this challenging problem.

Nevertheless, despite the gaps that remain in our understanding of primordial 
accretion flows, one point stands out clearly: the predicted accretion rates 
are very much larger than those inferred for local protostars, which are 
typically of the order of $10^{-4}$--$10^{-5} \: \msun \: {\rm yr}^{-1}$ for 
class 0 objects (see, e.g.\ \citealt{maret02}; \citealt{beuth02}), and which
decrease significantly as the protostar evolves \cite{andre00}. 

The reason for this difference is straightforward. On purely dimensional 
grounds, we would expect the time taken to accrete a mass $M$ of gas to be 
of the order of the free-fall timescale for the gas, unless some other 
effect, such as magnetic support or protostellar feedback, were to retard 
the collapse. Therefore, we can write the {\em mean} accretion rate for the 
gas as 
\begin{equation}
\dot{M} \propto \frac{M}{t_{\rm ff}},
\end{equation} 
where we expect the constant of 
proportionality to be of order unity. For gas with a mean density 
$\bar{\rho}$, we have $t_{\rm ff} \propto \bar{\rho}^{-1/2}$, and hence 
\begin{equation}
\dot{M} \propto M \bar{\rho}^{1/2}.
\end{equation} 
Now, if this mass of gas is close to 
being in hydrostatic equilibrium, which the results of \citet{abn02} show 
is a reasonable approximation for the gas surrounding the protostellar core, 
then $M$ must be of the order of the Jeans mass; if $M \ll M_{\rm J}$, the 
gas would not be collapsing, while if $M \gg M_{\rm J}$, it would almost 
certainly have fragmented. Therefore, the accretion rate scales as
\begin{equation}
\dot{M} \propto M_{\rm J} \; \bar{\rho}^{1/2}, 
\end{equation} 
or, since $M_{\rm J} \propto (T^{3}/ \bar{\rho})^{1/2}$, 
\begin{equation}
\dot{M} \propto T^{3/2}.
\end{equation}
Since the 
minimum temperature reached by the primordial gas
is more than an order of magnitude larger than the temperature characteristic
of local prestellar cores (which is typically of order $10 \: {\rm K}$; see 
\citealt{wak02}), we would expect the accretion rate
to be correspondingly greater, which is precisely what we find in the
detailed models discussed above.

\begin{figure}
\centering
\epsfig{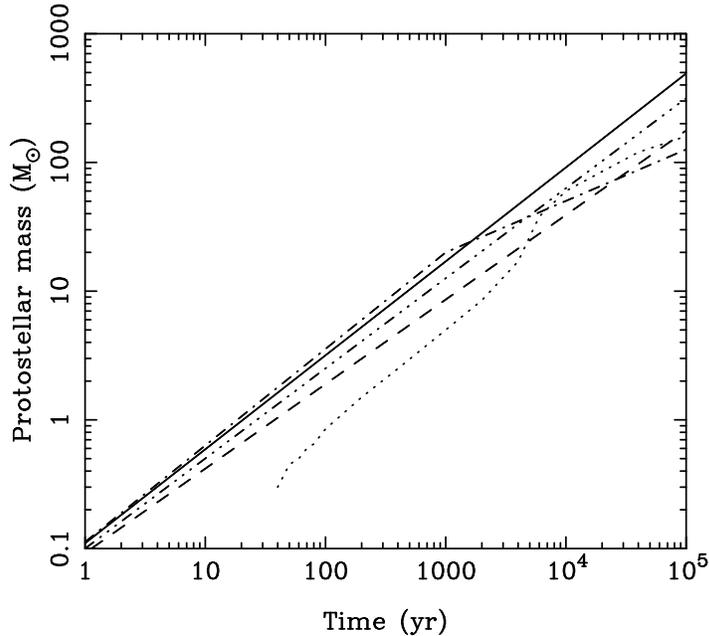}
\caption{The time-dependent protostellar masses produced by the 
accretion rates shown in  figure~\ref{accr_rate}.}
\label{final_mass}
\end{figure}

In the absence of significant protostellar feedback, these large predicted
accretion rates will lead to large final stellar masses. This is clearly 
demonstrated in figure~\ref{final_mass}, where I plot the final stellar 
mass as a function of time for all of the models discussed above. 
In every case, the final stellar mass grows to more than $100 \msun$ 
in less than $10^{5} \: {\rm yr}$. Therefore, unless feedback from the 
protostar can
significantly reduce the amount of material that the star accretes over 
its lifetime, it will inevitably become very massive, and will either
end its life as a pair-instability supernova (if its final mass lies in the
range $140 < M < 260 \: \msun$), or by collapsing directly to form a 
black-hole \cite{fwh01,hfwlh03}. The possible consequences of this are 
discussed in some detail by \citet{ybh04} and I will not discuss them here.

\subsection{Modeling protostellar feedback} 
Any potential form of feedback will be powered, directly or indirectly, from 
one of two sources: the energy released by the infalling matter, or the energy
produced by nuclear burning within the protostar. To model the former, we must
model gas flow near the surface of the protostar, paying particular attention 
to the properties of the accretion shock and the circumstellar accretion disk.
To model the latter, we must model the internal 
structure of the protostar. In practice, since the dominant energy source 
will change over time, from accretion at early times to nuclear burning
at late times, an ideal model should treat both regions, together with as 
much of the surrounding gas as possible.

Unfortunately, computational limitations again restrict us to more limited 
models, and we are forced to approximate. The most significant approximation
that is commonly made is the assumption of spherical symmetry. This is a 
reasonable approximation for the protostar itself, provided rotational 
effects are not significant, but it does not allow us to treat any processes 
involving the accretion disk, and is therefore rather limiting. On the other
hand, it dramatically reduces the computational requirements of the problem, 
and consequently continues to be a widely used approximation. Indeed, the 
only treatment of primordial protostellar structure and feedback of which I
am aware that does not assume spherical symmetry is the recent work of 
\citet{tm04} and \citet{tb04}, which is discussed in some detail later in this
section.

It is also common to further simplify the problem by splitting it into two
pieces, and considering the evolution of the structure of the protostar 
(which will strongly influence the strength of any feedback) separately 
from the effect of feedback on the flow. In other words, studies of 
protostellar structure generally assume a constant accretion rate, while
calculations focused on the effects of feedback on the accretion flow 
frequently assume a constant energy source. This separation is purely 
pragmatic; it is easier to study the different processes separately before
combining them in more realistic coupled models.

\subsubsection{The evolution of protostellar structure} 
The evolution in the structure of a primordial protostar as it accretes matter
from its surroundings was first studied in detail by \citet{sps86a}. Their
strategy followed that of \citeauthor{sst80a} (1980ab,~1981), who had 
previously studied a similar problem for the case of a low-mass, population I 
star. 

They assume that the accretion 
process can be treated as a series of quasi-steady-state accretion flows onto
a hydrostatic core, which is bounded by a strongly radiating accretion shock.
Within the core, the standard stellar structure equations are solved, with 
the assumption that deuterium burning is the only source of nuclear energy.
Outside the core, the treatment depends on the optical depth of the gas. If 
the gas is optically thin to the radiation from the accretion shock, then
the accretion flow is assumed to be in free-fall. Otherwise, a more detailed
calculation is made that incorporates the effects of the radiation force on
the infalling gas. The accretion shock itself is treated as a simple 
discontinuity; no attempt is made to model its structure in any detail.
Since the thickness of the shock is small compared to the size of the core, 
this should be a good approximation. 

\citet{sps86a} begin with an initial core mass of $0.01 \: \msun$,
and give the core an arbitrary initial distribution of specific entropy:
\begin{equation}
 s(M) = s_{0} + \beta \frac{k}{m_{\rm H}} \left(\frac{M}{\msun} \right)^{2},
\end{equation} 
where $\beta = 7.39$ and $s_{0}$ is calculated from the adopted central
temperature ($T_{\rm c} = 10^{5} \: {\rm K}$) and density ($\rho_{\rm c} = 
0.28 \: {\rm g} \: {\rm cm}^{-3}$) using an equation of state taken from 
\citet{eff73}. The outer boundary condition is fixed by the accretion rate,
which \citeauthor{sps86a} take to be constant, with a value 
$4.41 \times 10^{-3} \: \msun \: {\rm yr}^{-1}$.

Starting from these initial conditions, \citeauthor{sps86a} calculated the
subsequent evolution of the protostar until the core mass reached a value
of $10.5 \: \msun$. They found that the evolution could be divided into three
qualitatively distinct phases. 

In the first phase, which lasts until the core mass reaches $0.1 \: \msun$, 
the protostar relaxes from its initial entropy profile into one appropriate 
for the particular choice of accretion rate. This `decay of transients' phase
indicates that although the initial conditions are probably incorrect in 
detail, the flow soon loses all memory of them, and therefore any inaccuracy 
at this stage is unlikely to affect the later results.

Once the initial transients have died away, the protostar enters the second
phase of its evolution. During this phase, its central temperature remains
low ($T_{\rm c} \sim 10^{5} \: {\rm K}$), resulting in a high interior
opacity and hence a low interior luminosity. Consequently, the evolution of
the core during this phase is almost adiabatic; although the core continues 
to gradually contract, this contraction does not lead to any increase in the
central entropy. Since the postshock entropy increases over time due to the
increasing strength of the accretion shock (which is itself a natural result
of the increasing protostellar mass), the core develops an off-centre 
distribution of entropy and temperature. 

The gas surrounding the accretion shock remains optically thick throughout 
this period. This is a direct result of the high accretion rate, which 
produces a highly luminous accretion shock. This produces sufficient 
radiation to partially ionize the preshock gas in the vicinity of the
shock, creating a structure known as a radiative precursor. The $\Hm$ opacity
of the dense, partially ionized gas in this radiative precursor is more than 
sufficient to make it optically thick. \citeauthor{sps86a} show that the core 
radius during this period evolves as
\begin{equation}
 R_{*} = 48.1 \left(\frac{M_{*}}{\msun}\right)^{0.27} 
 \left(\frac{\dot{M}}{\dot{M}_{0}}\right)^{0.41} \: \rsun, 
\end{equation}
where $\dot{M}_{0} = 4.41 \times 10^{-3} \: \msun \: {\rm yr}^{-1}$, while
the photospheric radius evolves as
\begin{equation}
 R_{\rm p} = 66.8 \left(\frac{M_{*}}{\msun}\right)^{0.27} 
 \left(\frac{\dot{M}}{\dot{M}_{0}}\right)^{0.41} \: \rsun, 
\end{equation}
so $R_{\rm p} > R_{*}$ throughout. The strong $\Hm$ opacity also keeps the 
photospheric temperature low ($T_{\rm p} \sim 5000 \: {\rm K}$), which 
prevents the protostar from being able to ionize material outside of its
photosphere. 

This near-adiabatic accretion phase comes to an end once the cooling time of
the core, given approximately by the Kelvin-Helmholtz timescale
\begin{equation}
 t_{\rm KH} = \frac{G M_{*}^{2}}{R_{*} L_{*}},
\end{equation}
becomes comparable to the accretion timescale $t_{\rm acc} = M_{*} / \dot{M}$.
This occurs for a core mass $M \sim 1 \msun$, and results in the core
entering a phase of homologous collapse, while energy and entropy are 
transferred outwards in the form of a `luminosity wave'. The radial position
of the luminosity peak moves outwards towards the accretion shock, 
reaching it at about the time that the core mass has reached $8 \: \msun$. 
This results in a rapid swelling of the outermost layers, which weakens the 
accretion shock and leads to it becoming optically thin. \citeauthor{sps86a} 
terminate their simulation shortly afterwards, once the core mass has 
reached $10.5 \: \msun$.

Although \citeauthor{sps86a} include deuterium burning as a possible energy
source, in practice they find that it plays no role at this stage of the
protostar's evolution, as its central temperature remains too low to ignite
deuterium.
However, since the central temperature and density are both rising sharply
at the end of the simulation as the central regions of the core collapse
homologously, it is reasonable to assume that deuterium ignition will soon 
take place. \citet{sps86b} study the onset of deuterium burning and the
later onset of hydrogen burning in a subsequent simulation of the pre-main
sequence evolution of a $5 \: \msun$ primordial protostar. Their initial 
conditions are taken from \citet{sps86a}, but the accretion rate is now set
to zero. \citet{sps86b} find that deuterium ignites approximately 
$6000 \: {\rm yr}$ after the 
beginning of their simulation, with hydrogen ignition following after
$2 \times 10^{5} \: {\rm yr}$. The protostar eventually reaches the zero-age
main sequence approximately $10^{6} \: {\rm yr}$ into the simulation.

An alternative treatment of these later stages of evolution that does not
assume a negligible accretion rate is that of \citet{op01}. They construct 
their simulation in the same way as \citet{sps86a} and assume the same 
constant rate. The only significant technical differences between the
two simulations are that \citeauthor{op01} use zero metallicity opacities 
from \citet{lcs91} and \citet{ir96} in place of the older values used by 
\citeauthor{sps86a}, and that they begin their simulation at the start of the 
optically thin phase, when the core mass has already reached $M = 8 \: \msun$.\
However, unlike \citeauthor{sps86a}, they do not halt their simulation once 
the core mass reaches $10.5 \: \msun$; instead, they continue  until well 
after hydrogen ignition.

They find that the period of optically thin evolution identified by 
\citet{sps86a} lasts for only a short time; the core reaches a maximum 
radius of $220 \: \rsun$ for a core mass of $11.5 \: \msun$, but shortly 
afterwards begins a sustained process of contraction. The radiative precursor
reappears once the core mass reaches $12.4 \: \msun$ and persists for the
remainder of the simulation. As in the previous optically thick phase, 
$\Hm$ opacity keeps the photospheric temperature low. 

Within the core, deuterium burning begins once the mass of the core reaches 
$12 \: \msun$ (corresponding to a time $t = 1000 \: {\rm yr}$ after the 
beginning of the simulation, given the assumed accretion rate), and is 
complete by the time the mass has reached $30 \: \msun$ (corresponding to
$t = 5000 \: {\rm yr}$). It does not contribute significantly to the 
protostellar luminosity, and has little effect on the structure of the 
protostar.

Hydrogen ignition follows once the core mass reaches $80 \: \msun$ 
(corresponding to $t = 1.6 \times 10^{4} \: {\rm yr}$). At the same time,
the internal luminosity nears the Eddington value
\begin{eqnarray} 
 L_{\rm Edd} & = & \frac{4 \pi c G M m_{\rm p}}{\sigma_{T}} \\ 
 & = & 1.26 \times 10^{38} \left(\frac{M}{\msun}\right) \: \lsun,
\end{eqnarray}
triggering a second phase of expansion. The outer layer of the core moves
out from $10 \: \rsun$ to $100 \: \rsun$, although it remains well
within the photosphere, which has a radius of approximately 
$1000 \: \rsun$ at this time. As the core expands, the accretion 
luminosity falls and the radiation force on the outer layers of the core 
declines. It soon becomes too small to maintain the expansion, and so the core
begins to contract rapidly for a second time. From this point on, however,
nuclear burning makes a substantial and increasing contribution to the total 
protostellar luminosity, which soon reaches $L_{\rm Edd}$ for a second time.
This triggers another phase of radiation-driven expansion, which this time is
strong enough to halt the accretion. This occurs once the core mass has 
reached $M \sim 300 \: \msun$, and \citeauthor{op01} terminate their 
simulation at this point.

In order to assess the dependence of this result on the adopted accretion 
rate, \citet{op03} performed a similar analysis for a range of different 
values of $\dot{M}$. They defined a fiducial accretion rate 
$\dot{M}_{\rm fid} = 4.41 \times 10^{-3} \: \msun \: {\rm yr}^{-1}$ 
corresponding to the value adopted by \citet{sps86a} and \citet{op01}, and 
studied models with rates $\dot{M} = (0.25, 0.5, 1.0, 2.0) \times 
\dot{M}_{\rm fid}$, as well as a model using the time-dependent accretion
rate predicted by \citet{abn02}. 

The earliest stages of protostellar evolution are qualitatively the same in 
all of these models: we see the same sequence of adiabatic growth, 
propagation of a luminosity wave that triggers expansion of the outer layers, 
followed by rapid contraction. There {\em are} quantitative differences; for
instance, protostars with a larger $\dot{M}$ have a larger radius at a given
mass. However, significant differences in behaviour do not become apparent
until the end of the rapid contraction phase. In the fiducial case, 
\citet{op03} confirm their previous result: they find two episodes of 
radiation-driven expansion, the second of which is strong enough to terminate
accretion onto the protostar. In the $\dot{M} = 2 \dot{M}_{\rm fid}$ case,
however, they find that the initial phase of expansion is strong enough to
halt the accretion, thanks to the larger accretion luminosity associated with
the larger accretion rate. Consequently, the final protostellar mass is 
smaller, being approximately $90 \: \msun$. 

In the two models with $\dot{M} < \dot{M}_{\rm fid}$, however, the outcome
is rather different. Core contraction comes to an end shortly after the onset
of hydrogen burning, but there is no subsequent phase of radiation-driven
expansion, as the protostellar luminosity is never more than 70\% of 
$L_{\rm Edd}$. Instead, the core relaxes quickly onto the zero-age main 
sequence (ZAMS), continuing to accrete all the while.

\citet{op03} show that there is a critical accretion rate $\dot{M}_{\rm crit}$
that separates these two outcomes. Protostars with $\dot{M} < 
\dot{M}_{\rm crit}$ can reach the zero-age main sequence while still 
accreting, and can therefore grow to extremely large masses, while protostars 
with $\dot{M} > \dot{M}_{\rm crit}$ will undergo radiation-driven expansion 
before reaching the ZAMS, and will therefore have smaller masses. To evaluate
$\dot{M}_{\rm crit}$, \citeauthor{op03} equate the total luminosity of
a protostar that has just reached the ZAMS with the Eddington 
luminosity:
\begin{equation}
 L_{\rm Edd} = L_{\rm ZAMS} + \frac{GM_{*} \dot{M}_{\rm crit}}{R_{\rm ZAMS}},
\end{equation}
where the second term on the right-hand side represents the accretion 
luminosity. This equation can be rewritten as
\begin{equation}
 \dot{M}_{\rm crit} = \frac{4 \pi c R_{\rm ZAMS}}{\kappa_{\rm es}} \left(
 1 - \frac{L_{\rm ZAMS}}{L_{\rm Edd}} \right),
\end{equation}
where $\kappa_{\rm es}$ is the electron scattering opacity. Evaluating this,
we find that 
$ \dot{M}_{\rm crit} \simeq 4 \times 10^{-3} \: \msun \: {\rm yr}^{-1}$,
coincidentally close to $\dot{M}_{\rm fid}$. In principle, 
$\dot{M}_{\rm crit}$ has a dependence on the mass of the protostar, but in 
practice this dependence is weak and may be neglected.

The final model that \citet{op03} consider is one with a time-dependent 
accretion rate taken from \citet{abn02}. In this model, the accretion rate
is initially much larger than $\dot{M}_{\rm crit}$, but decreases with time,
and falls below $\dot{M}_{\rm crit}$ when the core mass reaches $95 \: \msun$.
This initial behaviour of this model is very similar to that of the model 
with $\dot{M} = \dot{M}_{\rm fid}$, but the two diverge during the rapid 
contraction phase; the time dependent model undergoes a very brief period of
radiation-driven expansion, but thereafter re-contracts, and relaxes onto the
zero-age main sequence, following which its evolution is indistinguishable 
from that of the other models with $\dot{M} < \dot{M}_{\rm crit}$.

An alternative view of the evolution of protostellar structure is presented
by \citet{tm04}. In contrast to previous authors, they do not assume spherical
symmetry, allowing them to treat the case of accretion via a circumstellar 
disk. \citeauthor{tm04} fix the size of the disk by assuming that angular 
momentum is conserved by gas within the sonic point of the accretion flow.
This allows them to write the disk radius as
\begin{equation}
 r_{\rm d} = f^{2}_{\rm Kep} \left(\frac{M_{\rm sp}}{m_{\rm *d}} \right) 
r_{\rm sp},
\end{equation}
where $r_{\rm sp}$ is the radius of the sonic point, $M_{\rm sp}$ is the mass
interior to the sonic point, $m_{\rm *d}$ is the mass interior to $r_{\rm d}$
(i.e.\ the mass of the protostar plus the disk) and $f_{\rm Kep}$ is the 
ratio of the rotational velocity of the gas to the Keplerian orbital velocity
$v_{\rm Kep} = (GM / r)^{1/2}$, evaluated at the sonic point. 
\citeauthor{tm04} fix $r_{\rm sp}$ using their analytic accretion flow 
solution, discussed in the previous section, and adopt 
$f_{\rm Kep} = 0.5$, based on the results of \citet{abn02}. They show that in
this case, the accretion disk radius becomes
\begin{equation}
  r_{\rm d} = 3.44 \left( \frac{f_{\rm Kep}}{0.5} \right)^{2} 
\epsilon^{-9/7}_{\rm *d} \left( \frac{m_{\rm *d}}{\msun} \right)^{9/7} 
K^{\prime-10/7} \: {\rm AU},
\end{equation}
where $K^{\prime}$ is given by Equation~(\ref{kprime}), and where 
$\epsilon_{\rm *d}$ is the fraction of the infalling gas that reaches the
disk or the star. In the absence of protostellar feedback, 
$\epsilon_{\rm *d} = 1$. 

Provided that $r_{\rm d} \gg r_{*}$, which will generally be the case, the 
bulk of the gas will accrete first onto the disk and only later onto the 
protostar. Therefore, the accretion rate onto the protostar, and hence its
structure, will be determined in large part by the behaviour of the disk.
To determine the disk structure, \citet{tm04} use the standard theory of 
steady, thin viscous accretion disks (as outlined in \citealt{ss73} 
or \citealt{fkr95}), with a spatially constant viscosity parameter $\alpha$. 
The dominant source of this viscosity and the appropriate value for 
$\alpha$ remain uncertain, much as they do in the analogous situation in 
present-day star formation. Possible sources of viscosity include 
gravitational instabilities within the disk 
\cite{lar84,lp87,b95,nm00,ga01,jg03}, tidal interactions with 
external mass concentrations \cite{spr87,lar90,lp93,blond00,lar02}, and 
turbulence generated by the magnetorotational instability (MRI; see 
\citealt{bh91},~\citeyear{bh98})

The last of these will only operate if a sufficiently strong magnetic field
is present in the disk. \citet{kuls97} showed that a very small seed field
could be produced via the Biermann battery mechanism \cite{bier50} during the 
collapse of the protogalactic gas, and \citet{tb04} show that although
this field is initially too small to drive the MRI, a dynamo process acting in
the disk will rapidly amplify the field, which soon becomes strong 
enough to drive the instability. In view of this, \citet{tm04} examine the
behaviour of a disk with $\alpha = 0.01$, the appropriate value for a disk
susceptible to MRI \cite{bh98}.

Having determined the disk structure and the rate at which gas accretes from 
the disk onto the protostar, \citet{tm04} then solve for the evolution of the
protostellar structure using a modified version of the analytic approach
developed by \citet{nhn95} and \citet{nhmy00}. In this approach, the 
protostellar radius is found
balancing the rate of accretion of energy with the rate of change of the total
protostellar energy. The internal structure of the protostar is not solved for
explicitly; instead, it is approximated as a polytrope, with the polytropic
index fixed by comparison with the results of \citet{sps86a} and \citet{op01}.

\citet{tm04} show that if an accretion disk is not present (i.e. if 
$f_{\rm Kep} = 0$), and if $\dot{M} = \dot{M}_{\rm fid}$, then
this model successfully reproduces the results of \citet{sps86a} and 
\citet{op01}. They 
also demonstrate that the presence of a disk has a relatively small effect 
on the evolution of the protostar. The protostellar radius tends to be 
somewhat larger than in the spherical accretion case, but the protostar 
still evolves through the same progression of adiabatic growth, terminated 
by the emergence of a luminosity wave, followed by rapid contraction that 
ends once the protostar reaches the zero-age main sequence. The major 
difference from the spherical case is in the behaviour of the photosphere.
Because most of the gas accretes onto the protostar via the disk, the gas
density is significantly reduced in regions near the protostar but out of 
the plane of the disk. Consequently, the optical depth of these regions is
also significantly reduced, with the result that the flow becomes optically
thin early in its evolution. For example, in the model with 
$f_{\rm Kep} = 0.5$, the photosphere vanishes once the protostellar mass 
reaches $1 \: \msun$ and does not subsequently reappear.
As we will see below, this may have a major influence on the effectiveness
of radiative feedback from the protostar.

\subsubsection{The effects of feedback}
In order for the protostar to substantially reduce the rate at which it 
accretes, it must be able to transfer a significant amount of energy and/or 
momentum to the infalling gas. A number of possible mechanisms that 
accomplish this have been suggested, which fall under two broad headings: 
{\em radiative feedback}, where radiation from the protostar (or the 
accretion disk) is responsible for transferring energy and momentum directly 
to the infalling gas, and {\em mechanical feedback}, where the protostar 
transfers energy and momentum to some form of outflow, which subsequently 
transfers it to the infalling material.

In local star-forming regions, the dominant mechanism is a form of radiative 
feedback: radiation pressure exerted on the infalling dust grains by the 
protostar results in a substantial momentum transfer to the gas and prevents
massive stars from forming, unless the accretion rate is very large 
\cite{wc87}. In dust-free primordial gas, however, this process is clearly 
inoperative, and we must examine other possibilities.

One obvious possibility is that radiation within the optically thick 
rotational and vibrational lines of $\mHt$ may exert sufficient pressure
to slow or stop the infall. However, this seems unlikely to be the case:
in their simulation of protostellar formation, \citet{ripa02} compute the
total opacity of the $\mHt$ lines and show that it is never more than 5\% 
of the electron scattering opacity, implying that the protostar would
have to radiate a total luminosity in the $\mHt$ lines that was 
many times larger than the Eddington luminosity for this effect to be 
dynamically significant. 

A more interesting possibility is that the buildup of an \hii region around 
the protostar may terminate the accretion. This idea was first discussed in
the context of present day star formation by \citet{ls71}, and was re-examined
in the context of primordial star formation by \citet{oi02}. The basic 
mechanism is straightforward: as the protostar ionizes the gas, it transfers
to it a considerable amount of thermal energy. If the \hii region can expand
to a radius at which this thermal energy exceeds the gravitational binding 
energy of the gas, then the ionized gas outside this radius will become
unbound from the central protostar, and little if any of it will ultimately 
be accreted. To assess the effectiveness of this mechanism, we need to answer 
two basic questions: one, does an \hii region actually form? And two, if an 
\hii region does form, can it expand sufficiently to unbind the gas, or will it
instead be confined to a small radius by the inflow?

The answer to the first of these questions will depend on the effective
temperature of the protostar. An isolated, massive metal-free star on the 
main sequence
will have an effective temperature of approximately $10^{5} \: {\rm K}$ 
\cite{coj00}, and will emit a substantial number of ionizing photons, so in 
this case it is clear that an \hii region will form. On the other hand, in the 
accretion models of \citet{sps86a} and \citeauthor{op01}~(2001,~2003)
discussed in the previous section, the protostar 
is hidden within a much larger photosphere, with an effective 
temperature of only $6000 \: {\rm K}$, and so no \hii region will form until
the photosphere vanishes at late times.

Regarding the second question, \citet{oi02} show that if the accretion flow
onto the protostar is steady and spherically symmetric, then an \hii region
can expand sufficiently to unbind the surrounding gas only if it is powered
by a flux of ionizing photons that exceeds a critical value, $Q_{\rm crit}$.
They demonstrate that in order to calculate $Q_{\rm crit}$
correctly, it is necessary to take into account an additional type of
radiative feedback -- the force arising due to radiation scattering within 
the \hii region. There are two main components to this force. One is due
to Thomson scattering, and will be negligible until the protostellar
luminosity approaches $L_{\rm Edd}$. The other comes from the momentum 
transfer that occurs during photoionization, and for a gas in photoionization 
equilibrium the force per unit mass is given by \cite{haeh95}
\begin{equation}
F_{\rm rad} = \frac{h\nu_{\rm ion}}{c} \alpha_{\rm B} \frac{n_{{\rm e}} 
 n_{\Hp} }{\rho},
\end{equation}
where $\alpha_{\rm B}$ is the case B recombination coefficient and 
$h\nu_{\rm ion} \simeq 13.6 \: {\rm eV}$ is the mean energy of an ionizing
photon. This radiative force acts to reduce the infall velocity within the
\hii region, which leads to an increase in the density of the ionized gas,
since the assumption of steady flow implies that $\rho \propto v^{-1}$, where
$v$ is the infall velocity. This increased density leads in turn to a higher 
recombination rate, which limits the expansion of the \hii region. 
\citeauthor{oi02} show that the net effect is to make $Q_{\rm crit}$ very 
large; they find a value
\begin{equation}
Q_{\rm crit} = 6.4 \times 10^{52} \left( \frac{R_{\rm in}}{10 \: \rsun} 
\right)^{-1} \left( \frac{M}{100 \: \msun} \right)^{2},
\end{equation}
where $R_{\rm in}$ is the inner radius of the \hii region. For reasonable 
values for the stellar parameters, this gives a value of $Q_{\rm crit}$ that
is about a hundred times larger than the actual ionizing flux, suggesting that
even if an \hii region forms, it will be unable to halt accretion onto the
protostar. It is worth noting, however, that the evolution of the \hii 
region is likely to be very sensitive to the density distribution near the 
protostar and it is not clear that this conclusion will still hold if we 
relax some of the simplifying assumptions made above. It would be instructive 
to redo this calculation using a more realistic dynamical model.

A final form of radiative feedback that has attracted serious consideration
is the scattering of Lyman-$\alpha$ photons by the infalling gas. Within the
\hii region, this is of only minor importance, but in the surrounding \hi 
gas, where the optical depth to Lyman-$\alpha$ scattering is much higher, 
it may be far more significant \cite{bd89,bith90,haeh95}. \citet{dk76}
argue that the radiation pressure exerted by the Lyman-$\alpha$ photons
will rapidly expel all of the \hi gas near the protostar, thereby limiting 
the final stellar mass to $10 \: \msun$ or less. On the other hand, 
\citet{oi02} contend that their treatment overestimates the density of 
Lyman-$\alpha$ photons and so overestimates the effectiveness of this
mechanism. More recently, \citet{tm03} have presented preliminary results
from a calculation of the effects of Lyman-$\alpha$ scattering that assumes
a rotating, axisymmetric inflow. Their results appear to support the 
\citet{dk76} picture,  although they quote
a larger mass limit of order $20 \: \msun$. However, full details of their
calculations have not yet been published, so it is not possible to assess the
strength of their argument.

The effects of feedback in the form of protostellar outflows have been less
well studied than the various radiative effects discussed above. In the case
of radiatively driven outflows, such as stellar winds from O type stars,
this neglect is easy to understand -- these outflows are 
driven primarily by the scattering of photons in the resonance lines of metal
ions, and thus grow substantially weaker as the metallicity declines 
\cite{kud02}. In primordial gas, outflows of this type must rely on Thomson
scattering and thus will only become significant if the protostellar luminosity
reaches $L_{\rm Edd}$, which, as we have seen, will only occur if the accretion
rate exceeds $\dot{M}_{\rm crit}$.

Meanwhile, bipolar outflows of the kind that are ubiquitous in local 
star-forming regions have attracted little study because they are widely 
believed to be hydromagnetic in nature (see, e.g.\ \citealt{mm99}) and 
protogalactic magnetic fields
were thought to be too weak to power them. However, in a recent paper, 
\citet{tb04} have argued that the initial protogalactic magnetic field will
undergo substantial amplification by a helical dynamo operating in the 
turbulent accretion disk surrounding the protostar, and may therefore
become strong enough to drive an outflow. For a reasonable choice of
parameters, they find that an outflow with mechanical luminosity
\begin{equation}
 L_{\rm mech} \simeq 100 \left(\frac{M}{\msun} \right) 
 \left(\frac{\dot{M}}{10^{-2} \: \msun \: {\rm yr}^{-1}} \right) \: \lsun
\end{equation}
can be produced. Although this value is substantially less than the radiative
luminosity of the protostar, the outflow is far more effective than the 
radiation at transferring momentum to the surrounding gas. \citeauthor{tb04}
estimate that it will begin to remove a significant quantity of gas from the
core once the protostellar mass exceeds $20 \: \msun$, and that as much as
50\% of the core mass may have been removed by the time that the protostellar
mass reaches $100 \: \msun$. The outflow will also alter the density structure 
of the core and should therefore be taken into account when assessing the
effectiveness of radiative feedback.

\subsection{Summary}
Although much work remains to be done on developing a detailed understanding 
of primordial protostellar accretion, several key points are already clear.
First, the mean accretion rate of a primordial protostar is much larger than
that of its present-day counterparts, simply as a result of the higher gas
temperature of the protostellar core. Second, a large quantity of gas is
available to be accreted by the protostar, since the very low efficiency of
fragmentation means that it does not have to compete with a large number of
other protostars for the available gas. Third, many of the forms of 
protostellar feedback that serve to limit the masses of protostars forming at 
the present day either do not operate in the primordial case, or operate with
a reduced effectiveness. At the same time, the higher accretion rate implies
a larger ram pressure of the infalling gas, making the job of halting the
accretion harder than at the present day. Fourth, those forms of feedback that
do seem to be effective (Lyman-$\alpha$ scattering, hydromagnetic outflows)
do not become so until late times, after the protostar has already accreted 
a substantial quantity of gas. 

Taken together, these points strongly suggest that the first stars will be
very massive. Indeed, if this basic picture is correct, it is difficult to
see how accretion could be terminated early enough to produce a solar mass 
star, since the predicted accretion rates discussed earlier suggest that
this mass of gas will build up in only 10--20~yr.

The major uncertainties that remain are easily summarized:
\begin{enumerate}
\item[(i)] Is our basic picture of a single protostar per core correct,
or does a second stage of fragmentation occur at late times, after the gas
has become optically thick? 
\item[(ii)] Does a dynamically significant disk form? If so, how does it 
evolve, and how does it affect accretion onto the protostar?
\item[(iii)] Is the effective temperature of the protostar 6000~K (as 
suggested by the models of \citeauthor{sps86a}, or \citeauthor{op03}) 
or $10^{5} \: {\rm K}$ (as suggested by the models of \citeauthor{tm04})?
\item[(iv)] Are there other possible forms of feedback that we haven't yet
considered?
\end{enumerate}
As with many of the open questions concerning primordial star formation, 
resolution of these issues is likely to require detailed numerical 
simulations with an adequate treatment of the effects of radiative transfer.

\section{Conclusion}
The past ten years have seen significant advances in our knowledge of many 
aspects of primordial star formation, from the large scale environment in
which it occurs to the very small scale structure of the first protostellar
core. Although a number of issues remain unresolved, a consensus now exists on
the broad outlines of the process. 

We expect the first stars to form in small, $\mHt$-cooled protogalaxies, 
with masses of $10^{5}$--$10^{6} \: \msun$, at redshifts $z = 30$--40. 
Fragmentation of the gas within these protogalaxies will be inefficient, 
contrary to previous expectations, and in the smallest protogalaxies only 
a single dense core will form, with a mass of a few hundred solar masses. 

This core will collapse without fragmenting further until it becomes optically
thick. Its subsequent evolution is not entirely certain, but the most probable
outcome is the formation of a single low-mass protostar near its centre. This
protostar will accrete gas rapidly from its surroundings, and will soon become
very large. Protostellar feedback may act to limit the accretion rate at late
times, in which case the final mass of the star will be similar to that of a
Galactic O type star; otherwise, the final stellar mass will be limited only
by the amount of gas available, and will be of the order of a few hundred 
solar masses. The first stars will therefore end their lives either exploding
as supernovae, or collapsing directly to form black holes. Either way, there
should be none left alive at the present day.

The future also holds great promise for the study of primordial star
formation. New facilities such as {\sc alma} and {\sc jwst} will for 
the first time allow us to probe the earliest epochs of star formation,
and may allow us to test observationally the picture that I have outlined
above (although the practical difficulties will remain formidable). 
Meanwhile, further increases in computing power will allow us to perform 
increasingly detailed simulations and should soon allow us to fill in many
of the gaps in our current understanding of the formation of the first 
stars.

\acknowledgements
The author acknowledges useful discussions on various aspects of primordial
star formation with T. Abel, P. Brand, V. Bromm, G. Bryan, M.-M. Mac~Low,
M. Norman, and R. Larson. He would also like to thank the referee, V. Bromm, 
for a careful reading of the manuscript and a number of suggestions which have
helped to improve its clarity. Financial support for this work was provided by 
NSF grants AST99-85392 and AST-0307793, and by NASA Education grant 
NAG5-13028. This research has made substantial use of NASA's Astrophysics 
Data System.

\theendnotes

\end{article}
\end{document}